\documentclass[aps,prc,nofootinbib]{revtex4}
\usepackage{eurosym}
\usepackage{amsmath,amssymb}
\usepackage{graphicx}
\usepackage{dcolumn}
\usepackage{bm}
\usepackage{color}
\usepackage{multirow}
\usepackage{float}
\usepackage{array}
\usepackage{dcolumn}
\usepackage{booktabs}
\usepackage[detect-all]{siunitx}
\usepackage{makecell}
\usepackage{xcolor}
\usepackage{yfonts}

\graphicspath{ {/Users/Nastia/Desktop/research_dis/papers_writing/xenes/nov_5} }

\newcolumntype{P}[1]{>{\centering\arraybackslash}p{#1}}
\begin{document}

\title{Study of light $\phi$-mesic nuclei with HAL QCD $\phi N$ interactions
 }
\author{I. Filikhin$^1$, R. Ya. Kezerashvili$^{2,3,4}$, and B. Vlahovic$^1$}
\affiliation{\mbox{$^{1}$North Carolina Central University, Durham, NC, USA} \\
$^{2}$New York City College of Technology, The City University of New York,
Brooklyn, NY, USA\\
$^{3}$The Graduate School and University Center, The City University of New
York, New York, NY, USA\\
$^{4}$Long Island University, Brooklyn, NY, USA}

\begin{abstract}
We explore the possible existence of light $\phi$-mesic nuclei using HAL QCD $\phi N$ interactions for the $^2S_{1/2}$ and $^4S_{3/2}$ channels.
Particulary, 
using the Faddeev formalism in configuration space, the $\phi NN$ system, and $^{9}_{\phi}$Be and $^{6}_{\phi\phi}$He nuclei within the framework of the three-body cluster model, 
are investigated. The $\phi\alpha$ effective potential, obtained through a folding procedure, involves the HAL QCD $\phi N$ interaction in the $^4S_{3/2}$
channel  which does not lead to a bound state of the  $\phi N$
pair while  the $\phi N$  $^2S_{1/2}$ channel yields the bound state as the $^3_\phi$H nucleus. The $^4S_{3/2}$ potential ensures that the folding procedure is appropriate because there are no open channels like  $\phi+N$ and $\phi +2N $  near or below the
 $\phi+ 4N$ threshold, and it 
utilizes different matter distributions of $^4$He proposed in the literature. The folding potential is approximated by the Woods-Saxon formula.  The mirror systems $\phi$+$\alpha$+$\alpha$ and $\phi$+$\phi$+$\alpha$ have energy ranges from 1-11 MeV and 3-10~MeV, respectively.
The predicted binding energies represent the minimal values for the hypothetical $\phi$ mesic nuclei $^{5}_{\phi}$He, 
$^{9}_{\phi}$Be and $^{6}_{\phi\phi}$He.  The phenomenological $\alpha\alpha$ and
$\phi\phi$ potentials are adopted from the literature.
\end{abstract}




\maketitle

\section{Introduction}

The interaction of the $\phi$ meson with the nucleon, nuclei, and the potential to form $\phi$-mesic nuclei is an intriguing topic in nuclear physics. The coupling of the $\phi$ meson to nucleons suggests that stable formation of $\phi$-mesic nuclei is likely, and theoretical models and experimental searches continue to explore this possibility \cite{Sekihara2010,Hayano2010,Paryev2017,Krein2018,Tolos2020,Proc2024}. Understanding $\phi N$ and $\phi$ meson nucleus interactions could offer valuable insights into the behavior of strange quarks in the nuclear medium and the formation of exotic nuclear states.

The $\phi$ meson is a vector meson with a mass of approximately 1020 MeV composed of a strange quark and anti-strange quark. The interaction between the $\phi$ meson and nucleons is typically modeled by effective hadronic field theory or meson exchange models.  This is a natural extension of the pion-nucleon and kaon-nucleon interactions, where the strange quark content of the $\phi$ meson leads to a more subtle interaction between the $\phi$ meson and the nucleon.

A $\phi$-mesic nucleus is a nuclear state in which one or more $\phi$ mesons are bound within a nucleus. $\phi$-mesic nuclei are strongly interacting exotic many-body systems. The nuclear medium can influence the properties of the $\phi$ meson, 
leading to a different behavior than in free space. 
Though the $\phi$ meson properties in vacuum, such as mass and width, are well known, it is not clear how these properties
will change once it is put in a dense environment such as nuclear matter.
This is due to the medium's influence, such as the potential screening effects and the changed density of the nucleus, which can alter the interaction dynamics between the $\phi$ meson and nucleons. To explore how the $\phi$ meson behaves at finite density has been the goal of several past and near future planning experiments. The key issues here are firstly whether $\phi$ meson is indeed bound to nuclei, secondly by how much is the bounding energy of states and what are the properties of such states, thirdly on the potential modifications of the $\phi$ meson in nuclear matter and their experimental detection \cite{Gubler2021}.
Theoretically, the properties of the $\phi$ meson in nuclear matter have been discussed based on hadronic models  \cite{Weise1998,Oset2001,Cabrera2003,Cabrera2017}, the QCD sum rules \cite{QCDF1992,QCDF1997,QCDF2015,QCDF2016,QCDF2022}, and the effective Lagrangian approach \cite{CoPL,Co17,Gubler2025}.

Studying the composite system from two nucleons and $\Lambda$-, $\Xi$-, or $\Omega$-hyperon has attracted intense research interest
in many theoretical works  \cite{Filikhin2000,GV15,GV2016,GVV16,FSV17,GV0,GV,Gibson2020,HiyamaXi2020,EF21,Zhang2022,GV2022,ESE2023}.
 Since the beginning of the new millennium many studies were devoted to composed system of $\bar K$-meson and two and three nucleons, see \cite{Akaishi2002,Yamazaki2002,Yamazaki2004,Shevchenko2017,RKezNNNK} and references herein, and also the system from two nucleons and $\phi$ meson \cite{BSS,Bel2008,Sofi,EA24,FKVPRD2024}.

The ALICE Collaboration experimental study \cite{ALICE2021} of correlations of $p-\phi\oplus \bar p-\phi$ pairs, measured in high-multiplicity $pp$ collisions at $\sqrt{s}= 13$ TeV indicates scattering length, which was interpreted as an attractive interaction. Latter Lyu et al. \cite{Lyu22} constructed the interaction between the $\phi$ meson and the nucleon based on the $(2+1)$-flavor lattice QCD simulations with nearly physical quark masses: HAL QCD potential in the spin 3/2 configuration is obtained from first-principles $(2+1)$-flavor lattice QCD simulations in a large space-time volume. However, $\phi N$ was not bound in the spin-3/2 channel. Recently, in Ref. \cite{Chizzali2024} a fit was made to the data on the correlation function by constraining the spin 3/2 interaction using the scattering length determined by lattice QCD simulations \cite{Lyu22}. As a result, evidence for a $\phi N$  bound state was found for the spin 1/2 channel. In this paper, we studying the lightest $\phi NN$ and $\phi\phi N$  $\phi$-mesic systems using HAL QCD interactions in 3/2 and 1/2 spin channel in the framework of the Faddeev equations in configuration space.

Generally speaking, there is no
universal method to handle $p$-shell systems such as $\Lambda$-, $\Xi$-, $\Omega$-, $\Sigma$-, $\eta$-nucleus, etc and, therefore, investigations are carried out using a variety of approaches such as the G-matrix theory,
variational methods, hyperspherical harmonics expansion method, resonating group method,
expansion in harmonic oscillator bases, and cluster models, just to mention a few of them.
For example, for a description of $p$-shell hypernuclei various theoretical approaches, e.g., the shell model \cite{Shell1,Shell2,Shell3,MShell1,MShell2,MShell3,MShell4}, \textit{ab initio} no-core shell model \cite{NoShell1,MShell5,Wirth,MShell6,MShell7}, a mean field model based on realistic 2-body baryon interactions \cite {Hiyama16}, and cluster models \cite{Motoba832,Motoba83,Motoba85,HiyamaCluster2001,HiyamaPPNP2009} were developed. 

It is well known that $\alpha$ clustering plays a crucial role in light nuclei. In addition to the light nuclei, the existence
of $\alpha$ clustering in the medium-mass nuclei and $\alpha$ matter was studied, see, for example, \cite{Okada2024,Clark2024}. 
There is a considerable number of  studies of $p$-shell hypernuclei such as $%
_{\Lambda \Lambda }^{6}$He, $_{\Xi \Xi }^{6}$He, $_{\Omega \Omega }^{6}$He,
$_{\Lambda }^{9}$Be, $_{\Xi }^{9}$Be within three-body cluster models \cite%
{Motoba832,Motoba83,Motoba85,HiyamaCluster2001,HiyamaPPNP2009,Fujiwara2004,Suslov2004,IgorF2004,Hiyama2012FBS,Wu2020,HiyamaRev2018}. One of the advantage to investigate $p$-shell nuclei which exhibit the cluster phenomena in the framework of the cluster model is that for description of a relative motion of the clusters one can apply the Faddeev or Faddeev-Yakubovsky equations \cite{Lazauskas2020} - the rigorous method for description of a few-body system. In this paper, we present a study of $\phi$-mesic nuclei $_{\phi \phi }^{6}$He and $_{\phi}^{9}$Be in the framework of the three-body cluster model as $\phi + \phi + \alpha$ and $\phi + \alpha +\alpha$ systems using the Faddeev equations in configuration space. While the $_{\phi }^{5}$He nucleus is investigated within the two-body $\phi + \alpha$ cluster model. Such an approach requires an effective $\phi\alpha$ potential.

This article has two objectives: i. study the possible formation of $\phi NN$ and $\phi\phi N$ bound $\phi$-mesic system using the HAL QCD $\phi N$ potentials in spin channels 3/2 and 1/2, respectively; ii. study the possible formation of $^{5}_{\phi}$He, $^{9}_{\phi}$Be, and $^{6}_{\phi\phi}$He $\phi$-mesic nuclei as the $\phi+\alpha$, $\phi+\alpha+\alpha$, and $\phi+\phi+\alpha$ system, respectively,  in the framework of a two- and three-particle cluster models by employing a $\phi\alpha$ interaction constructed based on the HAL QCD $\phi N$ potential in  3/2 spin channel. For three-particle systems this study is carried out in the framework of Faddeev equations in configuration space.

The paper is organized as follows. The $\phi N$, $NN$, $\alpha\alpha$, and $\phi\phi$ interaction potentials are discussed in Sec. \ref{Interactions}. In Sec. \ref{Fadd}, we present the Faddeev equations formalism in
configuration space for the description of a three-particle system when two particles are identical. In Sec \ref{Calculations}, we propose the $\phi\alpha$ potential obtained based on the Wood-Saxson fit of the folding of the HAL QCD $\phi N$ interaction 
and present results of
numerical calculations for $\phi NN$ and $\phi \phi N$ systems and $^{5}_{\phi}$He,
$^{9}_{\phi}$Be, and $^{6}_{\phi\phi}$He $\phi$-mesic nuclei. The concluding remarks follow in Sec. \ref{sec:6}.

\section{Interaction Potentials}
\label{Interactions}

A study of $\phi NN$ ($^{3}_\phi$H), $\phi\phi N$ ($^{3}_{\phi\phi}$H), $\phi$+$\alpha$ ($^{5}_\phi$He), $\phi$+$\alpha$+$\alpha$ ($^{6}_{\phi\phi}$Be), and $\phi$+$\phi$+$\alpha$ ($^{6}_{\phi\phi}$He) $\phi$-mesic systems within a nonrelativistic potential model requires $\phi N$, $NN$, $\phi\alpha$, $\alpha\alpha$, and $\phi\phi$ interaction potentials.

\textbf{$\phi N$ interaction.} The interaction of the $\phi$ meson with nucleons is a rich and ongoing area of research, bridging the study of hadron physics, nuclear physics, and quantum chromodynamics. Experiments, such as those conducted at facilities like J-PARC, the GSI, and the CERN SPS, have investigated the production of the $\phi $ meson in hadronic collisions, and $\phi$ meson photoproduction experiments conducted in JLab \cite{Strakovsky2020}.  The recent developments in both experimental and theoretical approaches gradually improve our understanding of this interaction. The interactions of the $\phi$ meson with nucleons have been analyzed within approuches such as chiral effective field theory, which includes both phenomenological and lattice QCD.

It has been suggested that the QCD van der Waals interaction, mediated by multi-gluon exchanges, is dominant when the interacting two color singlet hadrons have no common quarks
\cite{Brodsky1990}. Expecting the attractive QCD van der Waals force dominates the $\phi N$ interaction since the $\phi$ meson is almost a pure $s \bar s$ state following Brodsky, Schmidt, and de Teramond
\cite{Brodsky1990}, Gao, et al. \cite{G2001} suggested a Yukawa-type attractive potential:
\begin{equation}
V_{Y \phi N}(r)=
A\frac{e^{-\alpha r}}{r}, \;\; A=1.25 \; \text{GeV}, \alpha= 0.6 \; \text{fm}^{-1}.
\label{GaoPot}
\end{equation}%

The ALICE Collaboration reported values of the scattering length
and the effective range for the $\phi N$ interaction in the spin 3/2 configuration as $a^{(3/2)} = -1.43(23)^{+36}_{-06}$ fm and $r^{(3/2)}_{eff} = 2.36(10)^{+02}_{-48}$ fm \cite{ALICE2021}, showing that the interaction is attractive. In Ref. \cite{Lyu22}, the interaction between the $\phi $ meson and the
nucleon is studied based on the ($2+1$)-flavor lattice QCD simulations with
nearly physical quark masses.
The HAL QCD potential is obtained from first principles
(2 +1)-flavor lattice QCD simulations in a large spacetime
volume, $L^4 = $(8.1  \text{fm})$^4$, with the isospin-averaged masses of $\pi$,
$K$, $\phi$, and $N$ as 146, 525, 1048, and 954 MeV, respectively,
at a lattice spacing,
$a = 0.0846$ fm. Let us mention that such simulations together with the HAL
QCD method 
enable one to extract the $YN$ and $YY$
interactions with multiple strangeness, e.g., $\Lambda\Lambda$, $\Xi N$ \cite{Sasaki2020},
$\Omega N$  \cite{Iritani2019}, $\Omega\Omega$ \cite{OmegaOmega2018}, and $\Xi N$ \cite{HiyamaXi2020}. 
Using the HAL QCD method, 
the authors \cite{Lyu22} found that the $\phi N$ correlation
function is mostly dominated by the elastic scattering states in the $^{4}S_{3/2}$ channel without significant effects from the two-body $\Lambda
K(^{2}D_{3/2})$ and $\Sigma K(^{2}D_{3/2})$ and the three-body open channels
including $\phi N\rightarrow {\Sigma ^{\ast }K,\Lambda (1405)K}\rightarrow {%
\Lambda \pi K,\Sigma \pi K}$. The fit of the lattice QCD potential by the
sum of two Gaussian functions for an attractive
short-range part and a two-pion exchange tail at long distances with an
overall strength proportional to $m_{\pi }^{4n}$ \cite{Kreinm4}, has the
following functional form in the $^{4}S_{3/2}$ channel with the maximum spin $3/2$ \cite{Lyu22}:
\begin{equation}
V_{\phi N}^{3/2}(r)=\left(
a_{1}e^{-r^{2}/b_{1}^{2}}+a_{2}e^{-r^{2}/b_{2}^{2}}\right) +a_{3}m_{\pi }^{4}F(r,b_{3})\left( \frac{e^{-m_{\pi }r}}{%
r}\right) ^{2},  \label{HALQCD}
\end{equation}%
with the Argonne-type form factor \cite{Wiringa95}
\begin{equation}
F(r,b_{3})=(1-e^{-r^{2}/b_{3}^{2}})^{2}.  \label{Ffactor}
\end{equation}%
For comparison the lattice QCD $\phi N$ potential is also parameterized
using three Gaussian functions \cite{Lyu22}:
\begin{equation}
V_{G\phi N}(r)=\sum_{j=1}^{3}a_{j}\exp \left[ -\left( \frac{r}{b_{j}}\right)
^{2}\right]. \label{Gauss3}
\end{equation}%
Also, it was found that simple fitting functions, such as the Yukawa form (\ref{GaoPot}), cannot reproduce the lattice data \cite{Lyu22}.

Based on the findings in Refs. \cite{ALICE2021,Lyu22}, a fit was made to the data on the correlation function in Ref. \cite{Chizzali2024} by constraining the spin 3/2 interaction using the scattering length determined by lattice QCD simulations \cite{Lyu22}. As a result,
evidence for a $\phi N$ bound state was found for the spin 1/2 channel.
The HAL QCD potential in the $^{2}S_{1/2}$ channel with a minimum spin of $1/2$ \cite{Chizzali2024} has a much stronger attractive
$\beta$-enhanced short-range part 
and the same two-pion exchange
long-range tail as in the $^{4}S_{3/2}$ channel. The real part of the potential
in the $^{2}S_{1/2}$ channel reads \cite{Chizzali2024}
\begin{equation}
V_{\phi N}^{1/2}(r)=\beta \left(
a_{1}e^{-r^{2}/b_{1}^{2}}+a_{2}e^{-r^{2}/b_{2}^{2}}\right) +a_{3}m_{\pi
}^{4}F(r,b_{3})\left( \frac{e^{-m_{\pi }r}}{r}\right) ^{2},
\label{9}
\end{equation}%
where the factor $\beta =$6.9$_{-0.5}^{+0.9}$(stat.)$_{-0.1}^{+0.2}$(syst.).
The other values of the parameters are common in both $^{4}S_{3/2}$ and $%
^{2}S_{1/2}$ channels \cite{Chizzali2024}. The imaginary part of $\phi N$ potential related to
the 2nd-order kaon exchange and corresponds to absorption processes. A
proportionality coefficient for this part is $\gamma =$0.0$_{-3.6}^{+0.0}$%
(stat.)$_{-0.18}^{+0.0}$(syst.)
\cite{Chizzali2024}.
The potential $V^{1/2}_{\phi N}$ is five times deeper at $0 <r< 0.1$ fm and twice wider at distances $0.1 <r< 0.2$ fm 
than $V^{3/2}_{\phi N}$.

\begin{table}[!ht]
\begin{center}
\caption{The parameters for the $\phi N$ potential in the $^{4}S_{3/2}$ and $^{4}S_{3/2}$ channels. Statistical errors are quoted in parentheses. For the $a_{3}m_{\pi }^{4n}$ column, $n=1$ and $n=0$ for $V_{\phi N}^{3/2}$ and $V_{G\phi N}^{3/2}$, respectively \cite{Lyu22}. 
The parameters for the singlet and triplet $NN$ interaction for the MT potential \cite{Malfliet1969,MTcorr}, $\alpha\alpha$ potential \cite{AliBodmer}, and $\phi\phi$ interaction \cite{Sofi}.}
\label{tpp}
\begin{tabular}{ccccccc}
\hline
\noalign{\smallskip} & \multicolumn{5}{c}{${\phi N}$ potential in the $%
^{4}S_{3/2}$ channel \cite{Lyu22} and in the $^{2}S_{1/2}$ channel \cite%
{Chizzali2024}} &  \\ \hline
\noalign{\smallskip} & $a_{1},$ MeV & $a_{2},$ MeV & $a_{3}m_{\pi }^{4n},$
MeV fm$^{2n}$ & $b_{1},$ fm & $b_{2},$ fm & $b_{3},$ fm \\ \hline
\noalign{\smallskip} $V_{\phi N}^{3/2}$ & -371(27) & -119(39) & -97(14) &
0.13(1) & 0.30(5) & 0.63(4) \\
 $V_{\phi N}^{1/2}$ & -392.0 & -145.0 & -83.0 &
0.128 & 0.284& 0.582\\
$V_{G\phi N}$ & -371(19) & -50(35) & -31(53) & 0.15(3) & 0.66(61) & 1.09(41)
\\ \hline
\noalign{\smallskip} & \multicolumn{5}{c}{Singlet $^{1}S_{0}$ and triplet $%
^{3}S_{1}$ $NN$ potential \cite{Malfliet1969,MTcorr}} &  \\ \cline{2-6}
\noalign{\smallskip} & $I,J$ & $V_{r},$ MeV & $V_{a},$ MeV & $\mu _{1},$ fm$%
^{-1}$ & $\mu _{2},$ fm$^{-1}$ &  \\ \cline{2-6}
\noalign{\smallskip} & 1,0 & -521.959 & 1438.72 & 1.55 & 3.11 &  \\
& 0,1 & -626.885 & 1438.72 & 1.55 & 3.11 &  \\ \cline{2-6}
& \multicolumn{6}{c}{ABa $\alpha\alpha $ potential \cite{AliBodmer,FJ}} \\
\cline{2-6}
\noalign{\smallskip} & $l$ & $V_{r},$ MeV & $V_{a},$ MeV & $\mu _{1},$ fm & $%
\mu _{2},$ fm &  \\ \cline{2-6}
\noalign{\smallskip} & 0 & -30.18 & 125.0 & 2.85 & 1.53 &  \\ \cline{2-6}
& 2 & -30.18 & 20.0 & 2.85 & 1.53 &  \\ \cline{2-6}
& 4 & -130.0 &  & 2.11 &  &  \\ \cline{2-6}
& \multicolumn{6}{c}{$\phi \phi $ potential \cite{Sofi}} \\ \cline{2-6}
\noalign{\smallskip} &  & $V_{r},$ MeV & $V_{a},$ MeV & $\mu _{1},$ fm$^{-1}$
& $\mu _{2},$ fm$^{-1}$ &  \\ \cline{2-6}
\noalign{\smallskip} &  & 1000 & 1250 & 2.5 & 3 &  \\ \cline{2-6}
\end{tabular}
\end{center}
\end{table}

Although both (\ref{HALQCD}) and (\ref{Gauss3}) fits  provide an equally good result in reproducing the lattice data \cite{Lyu22}, below we perform calculations with both the lattice QCD $V_{\phi N}$ potential with a two-pion exchange tail and a purely phenomenological sum of three Gaussian,  $V_{G\phi N}$, potential. Parameters for these potentials are given in Table \ref{tpp}. Let us mention although the HAL QCD $\phi$-$N$ potential in $^{4}S_{3/2}$ channel with the maximal spin 3/2 is found to be attractive for all distances and reproduces a two-pion exchange tail at long distances \cite{ALICE2021,Lyu22}, no bound state is found with this interaction for $\phi N$ and $\phi NN$ systems \cite{FKVPRD2024}. Thus, the HAL QCD $\phi N$ potential in the
 $^{4}S_{3/2}$ channel does not provide enough attractiveness to support either the $\phi N$ \cite{Lyu22} or $\phi NN$ bound states \cite{FKVPRD2024}. 

\textbf{$NN$ interaction}.
We employ the same $NN$ MT-I-III potential \cite{Malfliet1969,MTcorr} as in \cite{BSS,Bel2008,Sofi,EA24}. The latter allows the comparison of the results. The singlet and triplet $NN$ interaction parameters for MT potential \cite{Malfliet1969,MTcorr} are given in Table \ref{tpp}.

\textbf{$\phi\alpha$ interaction.} Meson-nucleus systems bound by attractive interactions
are strongly interacting exotic many-body systems. The study of the  $^6_{\phi\phi}$He and $^9_\phi$Be nuclei within a three-body cluster model needs a $\phi\alpha$ potential. This potential might be approximated by “folding” the $\alpha$-particle nuclear density distribution with the $\phi N$ interaction.

In the single folding model the $\phi\alpha$ potential, $V_{\phi\alpha}^{F}(r)$, can be obtained as \cite{Satchler1983}:
\begin{equation}
\label{folding}
V_{\phi \alpha }^{F}(r)=\int \rho (\mathbf{x)}V_{\phi N}(\mathbf{r-x)}d\mathbf{x},
\end{equation}
where  $\rho (\mathbf{x)}$ is the density of nucleons in $^{4}$He and $\left\vert \mathbf{r-x}\right\vert $ is the distance between the $\phi$ meson and nucleon. Considering the central symmetry of the $V_ {\phi N}(r)$ potential and density $\rho(r)$,  expression (\ref{folding}) reads
\begin{equation}
\label{2Dintegral}
V_{\phi\alpha}^{F}(r)=4\int_{-1}^12\pi du\int_0^{\infty}dx \rho(x)V_{\phi N}(\sqrt{x^2+r^2-2xru})x^2.
\end{equation}
In Sec. IV using (\ref{2Dintegral}) and the mater density distribution in $^{4}$He we obtain a folding potential and construct a Wood-Saxon (WS) type interaction to simulate the $\phi$-$\alpha$ potential. We also simulated WS $\phi$$\alpha$ potential, using a potential obtained based on an effective Lagrangian approach that includes $K\bar{K}$ meson loops in the $\phi$ meson self-energy \cite{Co17}.

\textbf{$\alpha\alpha$ interaction.} The interaction between two $\alpha$ particles is expressed as a combination of nuclear and Coulomb components:

\begin{equation}
\label{AL}
V_{\alpha\alpha}(r) = V_n(r) + V_C(r).
\end{equation}
Generally there are two approaches for description of the nuclear part of (\ref{AL}). The $\alpha\alpha$ potential, which reproduces the observed
$\alpha$-$\alpha$ scattering phase shift and the ground state of $^{8}$Be, and by including the Pauli exclusion operator into the Hamiltonian of systems, see reviews \cite{Hiyama2012FBS,HiyamaRev2018} and references herein. 
A nuclear part of (\ref{AL}) is typically described using various phenomenological local potential models, such as the double Gaussian Ali-Bodmer potential \cite{AliBodmer}, Morse potential \cite{Morse1929}, or double Hulthen potential \cite{Hulthen2016}. 
Suggested over 60 years ago potential \cite{AliBodmer} 
has been widely used for calculations of nuclei binding energies in the framework of the cluster model, see, for example, Refs. \cite{Jibuti1978,KezNP1984,KezYad1992,Igor2000,fed1,suslov2004,suslov2005,fed2,fed3,Ishikawa1,Ishikawa2,Ishikawa3}. With its four
parameters chosen to fit scattering data in the leading states $l = 0$, 2, 4 of angular momentum up to 24 MeV, this interaction
consists of an $l$-dependent inner repulsive Gaussian term and
an $l$-independent outer attractive Gaussian term.
In our calculations, we adopt a four-parameter double Gaussian potential \cite{AliBodmer}:
\begin{equation}
V_n(r) = V_r e^{-\mu_r^2 r^2} - V_a e^{-\mu_a^2 r^2},
\label{alfaalfa}
\end{equation}
where $V_r$ and $V_a$ represent the strengths of the repulsive and attractive parts of the potential in MeV, 
respectively. $\mu_r$ and $\mu_a$ denote their corresponding inverse ranges in fm$^{-1}$. 

\textbf{$\phi\phi$ interaction.} We use the phenomenological
$\phi$-$\phi$ potential from Ref. \cite{Bel2008,Sofi} in the form of a sum of two Yukawa terms:
\begin{equation}
\label{pp} V_{\phi \phi}(r)=V_{1}\frac{e^{-\mu_{1} r} }{r}-V_{2}\frac{e^{-\mu_{2} r} } {r}.
\end{equation}
The parameters of this potential were fixed by the position and
width of the $f_{2}$(2010) resonance which has only one decay channel
into two $\phi$ mesons \cite{Bel2008,Sofi}.

\section{Theoretical formalism}

\label{Fadd}
The $\phi NN$ system, the description of the
$^{9}$Be$_{\phi}$ and $^{6}$He$_{\phi\phi}$ $\phi$-mesic nuclei as the
$\phi$+$\alpha$+$\alpha$ and $\phi$+$\phi$+$\alpha$ 
in a cluster model, represent three-particle systems.
The three-body problem can be solved in the framework of the Schr\"{o}dinger
equation or using the Faddeev approach in the momentum \cite{Fad,Fad1} or
configuration \cite{Noyes1968,Noyes1969,Gignoux1974,FM,K86} spaces. The Faddeev
equations in the configuration space have different form depending on the
type of particles and can be written for three nonidentical particles, 
three particles when two are identical, and 
three identical particles. 
The identical
particles have the same masses and quantum numbers. We consider the $\phi$+$N$+$N$, $\phi$+$\phi$+$N$, $\phi$+$\phi$+$\alpha$ and $\phi$+$\alpha$+$\alpha$ systems in the framework of the Faddeev formalism 
to seek possible bound states of the $^3_{\phi}$H, $^3_{\phi\phi}$H, $^6_{\phi\phi}$He and $^9_\phi$Be $\phi$-mesic nuclei. 

In the Faddeev
method in configuration space, alternatively, to the finding the wave
function of the three-body system using the Schr\"{o}dinger equation, the
total wave function is decomposed into three components \cite%
{Noyes1968,FM,K86}:
\begin{equation}
\Psi (\mathbf{x}_{1},\mathbf{y}_{1})=\Phi_{1}(\mathbf{x}_{1},\mathbf{y}%
_{1})+\Phi_{2}(\mathbf{x}_{2},\mathbf{y}_{2})+\Phi_{3}(\mathbf{x}_{3},%
\mathbf{y}_{3}).  \label{P}
\end{equation}%
Each Faddeev component corresponds to a separation of particles into
configurations $(kl)+i$, $i\neq k\neq l=1,2,3$. The Faddeev components are
related to its own set of the Jacobi coordinates ($\mathbf{x}_{i}$, $\mathbf{%
y}_{i}$), $i=1,2,3$. There are three sets of Jacobi coordinates. The total
wave function can be presented by the coordinates of one of the sets as is
shown in Eq. (\ref{P}) for the set $i=1$. The mass-scaled Jacobi coordinates
$\mathbf{x}_{i}$ and $\mathbf{y}_{i}$ are expressed via the particle
coordinates $\mathbf{r}_{i}$ and masses $m_{i}$ in the following form:
\begin{equation}
\mathbf{x}_{i}=\sqrt{\frac{2m_{k}m_{l}}{m_{k}+m_{l}}}(\mathbf{r}_{k}-\mathbf{%
r}_{l}),\qquad \mathbf{y}_{i}=\sqrt{\frac{2m_{i}(m_{k}+m_{l})}{%
m_{i}+m_{k}+m_{l}}}(\mathbf{r}_{i}-\frac{m_{k}\mathbf{r}_{k}+m_{l}\mathbf{r}%
_{l})}{m_{k}+m_{l}}).  \label{Jc}
\end{equation}%
In Eq. (\ref{P}), the components depend on the corresponding coordinate set
which are expressed in terms of the chosen set of mass-scaled Jacobi
coordinates. The orthogonal transformation between three different sets of
the Jacobi coordinates has the form:
\begin{equation}
\left(
\begin{array}{c}
\label{tran}\mathbf{x}_{i} \\
\mathbf{y}_{i}%
\end{array}%
\right) =\left(
\begin{array}{cc}
C_{ik} & S_{ik} \\
-S_{ik} & C_{ik}%
\end{array}%
\right) \left(
\begin{array}{c}
\mathbf{x}_{k} \\
\mathbf{y}_{k}%
\end{array}%
\right) ,\ \ C_{ik}^{2}+S_{ik}^{2}=1, \quad k\neq i,
\end{equation}%
where
\begin{equation*}
C_{ik}=-\sqrt{\frac{m_{i}m_{k}}{(M-m_{i})(M-m_{k})}},\quad S_{ik}=(-1)^{k-i}%
\mathrm{sign}(k-i)\sqrt{1-C_{ik}^{2}}.
\end{equation*}%
Here, $M$ is the total mass of the system. Let us definite the
transformation $h_{ik}(\mathbf{x},\mathbf{y})$ based on Eq. (\ref{tran}) as
\begin{equation}
h_{ik}(\mathbf{x},\mathbf{y})=\left(C_{ik} \mathbf{x}+ S_{ik}\mathbf{y},
-S_{ik}\mathbf{x}+ C_{ik}\mathbf{y} \right).  \label{Trans}
\end{equation}
The transformation (\ref{Trans}) allows to write the Faddeev equations in
compact form. The components $\Phi_i(\mathbf{x}_{i},\mathbf{y}_{i})$ satisfy
the Faddeev equations \cite{FM} and can be written in the coordinate
representation as:
\begin{equation}
(H_{0}+V_{i}(|C_{ik}\mathbf{x}|)-E)\Phi_i(\mathbf{x},\mathbf{y})=-V_{i}(|C_{ik}%
\mathbf{x}|)\sum_{l\neq i}\Phi_l(h_{il}(\mathbf{x},\mathbf{y})).  \label{e:1}
\end{equation}
In Eq. (\ref{e:1}), 
$H_{0}=-(\Delta _{\mathbf{x}}+\Delta _{\mathbf{y}})$ is the
kinetic energy operator with $\hbar ^{2}=1$ and $V_{i}(|\mathbf{x}|)$ is the
interaction potential between the pair of particles $(kl)$, where $k,l\neq i$.
The
spin-isospin variables of the system can be represented by the
corresponding basis elements. After the separation of the variables, one can
define  the coordinate part the $\Psi^{R}$ of the wave function $\Psi =\xi
_{spin}\otimes \eta _{isospin}\otimes \Psi ^{R}$.

\begin{figure}[ht]
\begin{center}
\includegraphics[width=15pc]{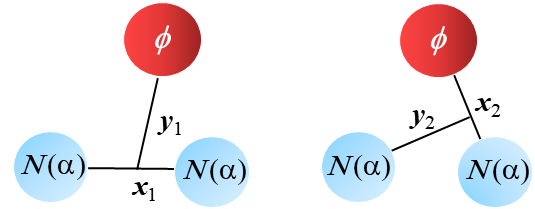}
\end{center}
\caption{Schematics for Jacobi coordinates for the $\phi$+$N$+$N$ ($\phi$+$\alpha$+$\alpha$) system. The coordinates correspond to two rearrangements: $(NN) \phi$ (($\alpha\alpha$)$\phi$) and $N(N\phi)$ ($\alpha(\alpha\phi)$).
}
\label{fig:03}
\end{figure}

The system of equations, Eqs. (\ref{e:1}), formulated for three nonidentical particles can be simplified in the case of two identical particles \cite{Kez2017,Kez2018PL,KezPRD2020}. The systems $\phi$ + $N$ + $N$ and $\phi$ + $\phi$ + $N$ are examples of such cases, where two identical nucleons or two identical $\phi$ mesons are present.
Correspondingly, within the cluster model, the $^{9}_{\phi}$Be and $^{6}_{\phi\phi}$He $\phi$-mesic nuclei can be treated as three-particle systems with two identical $\alpha$ particles or two $\phi$ mesons, respectively. The total wave function of these systems is given by:
\begin{equation}
\Psi =\Phi_1+\Phi_2\pm P\Phi_2. \label{e:0}
\end{equation}
Here, $P$ is the permutation operator for identical particles, which acts on the corresponding coordinates and spin-isospin variables of the Faddeev components. For two identical bosons, the sign is '+', whereas for two identical fermions, the sign '-' must be chosen.
Schematics picture for Jacobi coordinates in the $\phi$+$N$+$N$ ($\phi$+$\alpha$+$\alpha$) system is shown in Fig. \ref{fig:03}.  The coordinates correspond to two rearrangements: $(NN) \phi$ (($\alpha\alpha$)$\phi$) and $\phi(NN)$ ($\phi(\alpha$)$\alpha)$). For the $\phi$+$N$+$N$ system the wavefunction is antisymmetrized with respect to  nucleons and for the $\phi$+$\alpha$+$\alpha$ system the wavefunction is symmetrized with respect to $\alpha$ particles. For the $\phi$+$\phi$+$N$ system ($\phi$+$\phi$+$\alpha$)
one could replace the nucleons ($\alpha$-particles) by two $\phi$ mesons and the $\phi$ meson with the nucleon ($\alpha$-particle).
The Coulomb interaction between $\alpha$ particles is included as a perturbation on the left-hand side of Eq. (\ref{e:1}). As an example, Ref. \cite{KezJPG2024} presents the Faddeev equations for two identical particles while accounting for the Coulomb potential.
Below, we apply an $s$-wave model to the $\phi$ + $N$ + $N$ ($\phi$ + $\phi$ + $N$) and $\phi$ + $\phi$ + $\alpha$ systems. However, the treatment of the $\phi$ + $\alpha$ + $\alpha$ system must incorporate the first three orbital components of the $\alpha$–$\alpha$ potential, corresponding to $l = 0, 2, 4$.
Further details on our methods for solving the differential Faddeev equations (\ref{e:1}) for such systems can be found in \cite{KezPRD2020,KezJPG2024,Kez2017}.

\section{Results and discussion}
\label{Calculations}
\subsection{$\phi NN$ system}
The $\phi$-mesic $\phi NN$ system is considered in the framework of Faddeev equations in 
\cite{BSS}, using the variational folding method \cite{Bel2008}, and a two-variable integro-differential equation describing bound
systems of unequal mass particles \cite{Sofi}. Calculations were employed $\phi N$ potential from \cite{G2001}. 
The binding energy (BE) of $\phi d$ $\phi$-mesic nucleus was calculated by employing HAL QCD potential \cite{Lyu22} using the Schrodinger equation for Faddeev components 
expanded in terms of hyperspherical functions \cite{EA24}. The BEs reported in Refs. \cite{BSS,Bel2008,Sofi,EA24} are in the range of $\sim 6-39$ MeV.

Motivated by 
newly suggested HAL QCD potentials in the $^{2}S_{1/2}$ and $^{4}S_{3/2}$ channels with a minimal and maximal spin, respectively, we present calculations for the BEs for the $\phi N$ and $\phi NN$ in the framework of the Faddeev equations in configuration space. We compare our results with 
calculations  \cite{BSS,Bel2008,Sofi,EA24} as well.

Below we present the results of calculations for the feasibility of expected bound states for $\phi N$ and $\phi NN$ systems. For calculations of the BEs of these systems, we use the HAL QCD $\phi N$  potential in the $^{4}S_{3/2}$ and  $^{2}S_{1/2}$ channels  with the maximum and minimum spins, respectively, 
and employ the same $NN$ MT-I-III potential \cite{Malfliet1969,MTcorr} as in \cite{BSS,Bel2008,Sofi,EA24} for the comparison of the results. The input parameters for potentials are listed in Table \ref{tpp}. For comparison, we also perform BE calculations  for $\phi N$ and $\phi NN$ systems with previously suggested Yukawa-type $\phi N$ potential with parameters from \cite{G2001} and \cite{ALICE2021}.

The spin-isospin configurations of the $\phi NN$ system are
including four configurations:  the three are corresponding to states with   $S=0,1,2$ and $T=0$ and the fourth is the states with  $S=1$,  $T=1$ \cite{FKVPRD2024}.
The isospin state $T=0$ means that the considered system includes the deuteron ( $s=1$, $t=0$) with spin triplet nucleon-nucleon potential. The state $S=1$, $T=1$ has spin singlet nucleon-nucleon potential.

For calculations for the $S=1, T=0$ and $S=1, T=1$ states we used an averaged over spin-isospin variables potential.
Following \cite{Chizzali2024},  the  effective $\phi N$ potential, the averaged by $^{4}S_{3/2}$ and  $^{2}S_{1/2}$ channel potentials, is defined as
\begin{equation}
\label{Mix}
\bar V_{\phi N}=\frac{1}{3}V_{\phi N}^{1/2} + \frac{2}{3}V_{\phi N}^{3/2},\   S=1, T=0, \quad
\bar V_{\phi N}=\frac{2}{3}V_{\phi N}^{1/2} + \frac{1}{3}V_{\phi N}^{3/2},  \  S=1, T=1.
\end{equation}
\begin{table}[h!]
\caption{The scattering length $a_{\phi N}$ and  effective radius $r_{\phi
N}$  in fm and binding energies  $B_{2}^{\phi N}$  in MeV for $\phi N$ with the HAL QCD interactions in the $^{2}S_{1/2}$ and $^{4}S_{3/2}$ channels and mixed spin states, respectively. $B_{2}^{N N}$  in MeV is the $NN$ binding energy. $B_{3}^{\phi NN}$ in MeV is the binding energy of the $\phi d$ or $\phi NN$ in the
 spin-isospin states 
$(S,T)$.   The "UNB" indicates that no bound state is found. $B^{\phi
NN}_{3}$($V_{NN}=0)$ is the BE of the $\phi NN$ system when 
$NN$ interaction is omitted,  $V_{NN}=0$. 
}
\label{RtR3}
\begin{tabular}{cccccccc}
\hline
 \noalign{\smallskip}
$\phi N$ potential &$(S,T)$ & $a_{\phi N}$  & $r_{\phi
N}$ & $B^{\phi
N}_{2}$& $B^{NN}_{2}$&$B^{\phi NN}_{3}$&$B^{\phi
NN}_{3}$($V_{NN}=0$) \\
 \noalign{\smallskip}\hline \noalign{\smallskip}
$V_{\phi N}^{3/2}$ \cite{Lyu22}&($2,0$)   & -1.37  & 2.42 &UNB &2.23&UNB& UNB\\
$(\frac{1}{3}V_{\phi N}^{1/2}+\frac{2}{3}V_{\phi N}^{3/2})$\cite{Chizzali2024} &($1,0$)
& &   & UNB&2.23 &4.23& UNB\\
$(\frac{2}{3}V_{\phi N}^{1/2}+\frac{1}{3}V_{\phi N}^{3/2})$& ($1,1$)
&  &   & 0.47&UNB&6.82$^*$ &1.745\\
$V_{\phi N}^{1/2}$\cite{Chizzali2024}&($0,0$) & 1.5 & $\sim 0$ &  27.7&2.23  &64.17& 68.31\\
\noalign{\smallskip}\hline \noalign{\smallskip}
($V_{\phi N}^{3/2}$)x10.9/3 &($2,0$)   &   & &27.9 &2.23 &81.88& 61.05\\
\noalign{\smallskip}\hline \noalign{\smallskip}
\end{tabular}\\
$^*${\small Recalculated and corrected the value originally presented in Ref. \cite{FKVPRD2024}} \ \  \ \  \  \ \  \   
\end{table}
In Tables \ref{RtR3} and \ref{Rt33} we present the numerical results for the $\phi NN$ system obtained with the HAL QCD interactions and a Yukawa-type potential with a parameterization from \cite{G2001}.  The Yukawa-type potential with the parameterization from \cite{ALICE2021} gives  unbound $\phi N$ and $\phi NN$.
The results for the scattering length and effective radius with the potential in the $^{4}S_{3/2}$ channel are similar to the original one reported in  Ref. \cite{Lyu22} where the following value were obtained for scattering length $a_{\phi N}=-1.43(23)$$_{stat}(^{+36}_{-06})_{syst}$ and  effective radius $r_{\phi
N}=2.36(10)_{stat}(^{+02}_{-48})_{syst}$.

One can evaluate  the mass polarization term using the definition \cite{H2002,FG2002,F2018}:
\begin{equation}
\label{eq:delta}
\Delta =B^{\phi NN}_3(V_{NN}=0)-2B^{\phi N}_2.
\end{equation}
In Eq. (\ref{eq:delta}), $B^{\phi N}_2$ represents the $\phi N$ two-body binding energy, and $B_3(V_{NN}=0)$ denotes the three-body binding energy when the ${NN}$ interaction between identical particles is neglected, \textit{i. e.} $V_{NN}=0$. Generally, the value of $\Delta$ is positive and depends on the mass ratio of the particles (mainely) and the potentials. For the $\phi N$ averaged potential in the state ($1,1$), the mass polarization, $\Delta/B^{\phi NN}_3(V_{NN}=0)$, is about 50\% for a weakly attractive $NN$ potential. This value is comparable to those obtained for three-body systems with a mass ratio of close to 1. For example, calculations of the mass polarization for the $\bar{K}KK$ system yield values in the range of 24\%-30\% for different $\bar{K}K$ potentials \cite{KezPRD2020}. For the $\phi NN$ system in the state ($0,0$)  for this ratio, we obtained a value above 20\%. Note that the $\phi N$ potentials differ for the states ($1,1$) and ($0,0$).

Our calculations with the HAL QCD $\phi N$ interaction in the $^{4}S_{3/2}$ channel 
for the $\phi \phi N$ system did not reveal a bound state. In contrast, the HAL QCD $\phi N$ potential in the $^{2}S_{1/2}$ channel yields a deeply bound state with a binding energy of 179.3 MeV and when the interaction between two $\phi$ mesons is omitted, $V_{\phi\phi}=0$, the BE is 71.5 Mev. 

Generally, the latter potential exhibits extrimly strong attraction. Adding a single nucleon to the bound $\phi N$ system increases the $\phi NN$ binding energy by approximately 40 MeV due to common attractive of $NN$ (deuteron) pair and the $\phi N$ potential. To roughly estimate the binding energy of the $\phi + \alpha$ system, we might add twice  the binding energy of the $\phi NN$ system.  However, the resulting value of about 112 MeV $=140-28$ MeV for the $\phi + \alpha$ system is unphysically high. We understand that this estimation method is too simplistic and does not accurately capture the complex many-body interactions within the $\phi + \alpha$ system. Consideration of the alpha particle's structure requires a more sophisticated calculation. 

One can compare the binding energies of  the $\phi NN$ different states for the two scenarios: the full model and the restricted model with $V_{NN}=0$. The state ($1,1$) exhibits a regular relation, $B^{\phi NN}_3 \geq B^{\phi NN}_3(V_{NN}=0)$, which is typically true for systems where identical particles interact via an attractive potential. However, the state ($0,0$) shows the opposite relation when a strongly attractive $NN$ potential is present. A similar situation arises when identical particles experience repulsion. 
 \begin{table}[ht]
\caption{The binding energies  $B_{2}^{\phi N}$  in MeV for $\phi N$ in the $^{4}S_{3/2}$ channel, 
$B_{3}^{\phi NN}$ ($B_{3}^{\phi \phi N}$) in MeV is the BE of the $\phi NN$ ($\phi \phi N$) system in the spin-isospin states $(S,T)$ or isospin state of $NN$ pair, $t_{NN}$ .  
The "UNB" indicates that no bound state is found. $B_{3}^{\phi NN}(V_{NN}=0)$ ($B_{3}^{\phi \phi N}(V_{\phi \phi }=0)$) in MeV correspons to the calculations of BEs of the system $\phi NN$ ($\phi\phi N$) when $NN$ ($\phi\phi$) interaction is omitted. 
Here, we note the MT-I-III $NN$ potential as MT$^a$ from \cite{Malfliet1969}  and MT$^b$ from \cite{MTcorr}. }
\label{Rt33}
\begin{tabular}{ccccccc}
\hline
$\phi N$ & $a_{3}m_{\pi }^{4}$ & $NN$& $(S,T)$ & $B_{2}^{\phi N}$
& \multicolumn{2}{c}{$B_{3}^{\phi NN}$} \\ \hline
 & $-31$ & &  & UNB & \multicolumn{2}{c}{UNB} \\ 
$V_{G\phi N}^{3/2}$ & $-31$ & MT$^{b}$  & $(2,0)$ & UNB \cite{EA24}& \multicolumn{2}{c}{
6.7 \cite{EA24}} \\ 
 & $-56.44$ & &  & UNB & \multicolumn{2}{c}{6.77} \\ 
\hline
 & $-97$ &   &  & UNB & \multicolumn{2}{c}{UNB} \\ 
$V_{\phi N}^{3/2}$ & $-97$ &MT$^{b}$ &$(2,0)$  & UNB\cite{EA24} & \multicolumn{2}{c}{
6.8 \cite{EA24}} \\ 
 & $-167.6$ &  &  & UNB & \multicolumn{2}{c}{6.7} \\ 
\hline\hline
$\phi N$ & \multicolumn{2}{c}{$NN$} & $t_{NN}$ & $B_{2}^{\phi N}$ & $%
B_{3}^{\phi NN}$ & $B_{3}^{\phi NN}(V_{NN}=0)$ \\ \hline
 & \multicolumn{2}{c}{MT$^{a}$} &  & 9.43 & 37.9 
\cite{BSS} & -- \\ 
 & \multicolumn{2}{c}{MT$^{a}$} &  & 9.47 & 39.842 \cite%
{Sofi} & -- \\ 
$V_{Y\phi N}$ & \multicolumn{2}{c}{MT$^{a}$} & $1$& 9.43 & 38.51 & 25.26
\\ 
 & \multicolumn{2}{c}{MT$^{b}$} &  & 9.43 & 38.05 & 25.26
\\ 
 & \multicolumn{2}{c}{MT$^{b}$} &  & -- & 40 \cite{EA24} & --
\\ \hline
& \multicolumn{2}{c}{MT$^{a}$} & & 9.43 & 21.8 
\cite{BSS} & -- \\ 
 & \multicolumn{2}{c}{MT$^{a}$} &  & 9.47 & 23.609 \cite%
{Sofi} & -- \\ 
$V_{Y\phi N}$ & \multicolumn{2}{c}{MT$^{a}$} & $0$  & 9.43 & 22.61 & 25.26
\\ 
 & \multicolumn{2}{c}{MT$^{b}$} &  & 9.43 & 22.43 & 25.26
\\ 
 & \multicolumn{2}{c}{MT$^{b}$} &  &--  &23 \cite{EA24}  & --
\\ \hline\hline
$\phi N$& \multicolumn{2}{c}{$\phi \phi $} & $B_{2}^{\phi N}$ & $%
B_{3}^{\phi \phi N}$ & $B_{3}^{\phi \phi N}(V_{\phi \phi }=0)$ \\ \hline
$V_{Y\phi N}$ & \multicolumn{2}{c}{} &  9.43 & 77 \cite{BSS} & --
\\ 
$V_{Y\phi N}$ & \multicolumn{2}{c}{$V_{\phi \phi }$ \cite{Sofi}}  & 9.43 & 
77.73 & 26.90 \\ 
$V_{\phi N}^{3/2}$ & \multicolumn{2}{c}{}  & UNB & UNB & -- \\ 
$V_{\phi N}^{1/2}$ & \multicolumn{2}{c}{} & 27.7 & 179.3 & 
71.45 \\ \hline
\end{tabular}
\end{table}

To investigate this further, we performed additional calculations for the ($2,0$) state using a scaled $V_{\phi N}^{3/2}$ potential with a scaling factor $10.9/3$. This scaling factor leads to the BE 27.9 MeV for the $\phi N$ comparable with one obtained with $V_{\phi N}^{1/2}$ potential. The corresponding results are presented in the last row in Table~\ref{RtR3}. These results correspond to the case where the interaction between identical particles is attractive.
Our explanation for this discrepancy is that the potential $V_{\phi N}^{1/2}$ exhibits stronger attraction at short distances. As a result, the nucleons come very close to one another, which activates the repulsive core of the $NN$ interaction. This repulsion effectively weakens the attractive component of the $NN$ pair in this compact three-body system and leads to $B_{3}^{\phi NN}(V_{NN}=0)> B_{3}^{\phi NN}$. 

We compare the potentials presented in Table~\ref{Rt33} to demonstrate their dependence on the distance between the $\phi$ meson and the nucleon. The graphical results are shown in Fig.~\ref{f:1}. Of particular interest is the comparison between the $V_{\phi N}^{1/2}$ and the scaled $V_{\phi N}^{3/2}$ cases. The former represents a very short-range, deep potential, while the latter corresponds to a medium-range potential.
In terms of a square quantum well terminology, the $V_{\phi N}^{1/2}$ potential can be described as a narrow and deep potential, whereas the scaled $V_{\phi N}^{3/2}$ potential is a wider and shallower one.

The root-mean-square ($rms$)  distance between the $\phi$-meson and a nucleon in the $\phi N$ system is approximately 1.07 fm for the $V_{\phi N}^{1/2}$ potential, while the scaled $V_{\phi N}^{3/2}$ potential yields 1.32 fm. In $\phi NN$ calculations using the $V_{\phi N}^{1/2}$ potential, the $rms$ distance is 1.2 fm in the $\phi N$ subsystem and 1.5 fm in the $NN$ subsystem. These results highlight the compactness of the $\phi NN$ system compared to the deuteron, which has an rms distance of about 4 fm.
 The corresponding $rms$ distances between $\phi$ meson and nucleon  and between nucleons in $\phi N$ and $N N$ pairs are depicted in Figs. \ref{f:7}($a$) and ($b$). In Figs. \ref{f:7}($c$) and \ref{f:7}($d$) are shown the distances for the full $\phi N N$ model when thee particles interact with each other and the restricted model when the interaction between nucleons are omitted, $V_{NN} = 0$.
 
It has to be noted here that in the framework of the Faddeev equations formalism, the mean square root distance between each pair of three particles is evaluated as an averaged square of the Jakobi coordinate $\mathbf{x}_{\text{i}}$:
\begin{equation}
d_{i}=\sqrt{\int \Psi ^{\ast }(\mathbf{x}_{\text{i}},\mathbf{y}_{\text{i}})%
\mathbf{x}_{\text{i}}^{2}\Psi (\mathbf{x}_{\text{i}},\mathbf{y}_{\text{i}})d%
\mathbf{x}_{\text{i}}d\mathbf{y}_{\text{i}}}, \quad i=1,2,3.
\label{RK}
\end{equation}
In Eq. (\ref{RK}), the Jacobi coordinates $(\mathbf{x}_{\text{i}},\mathbf{y}_{\text{i}})$  defined without the mass scaling of Eq. ({\ref{Jc}). The wave function $\Psi (\mathbf{x}_{\text{i}},\mathbf{y}_{\text{i}})$ is normalized to 1.

Note that the $V_{\phi N}^{1/2}$ potential applied for  $\phi N N$ system leads to the $rms$ distance in the subsystem $\phi N $
of 1.2 fm and in the subsystem $N N$ of 1.5 fm. The last result show the compactness of the system due to the deuteron $rms$ distance is about 4 fm. 
The corresponding $rms$ distances between nucleons and $\phi$ meson and nucleon are depicted in Fig. \ref{f:7}. In Figs. \ref{f:7}($c$) and \ref{f:7}($d$) are shown the distances for the full model when three particles interact with each other and the restricted model when the interaction between nucleons is omitted, $V_{NN} = 0$. 

\begin{figure}[t]
\begin{center}
\includegraphics[width=22pc]{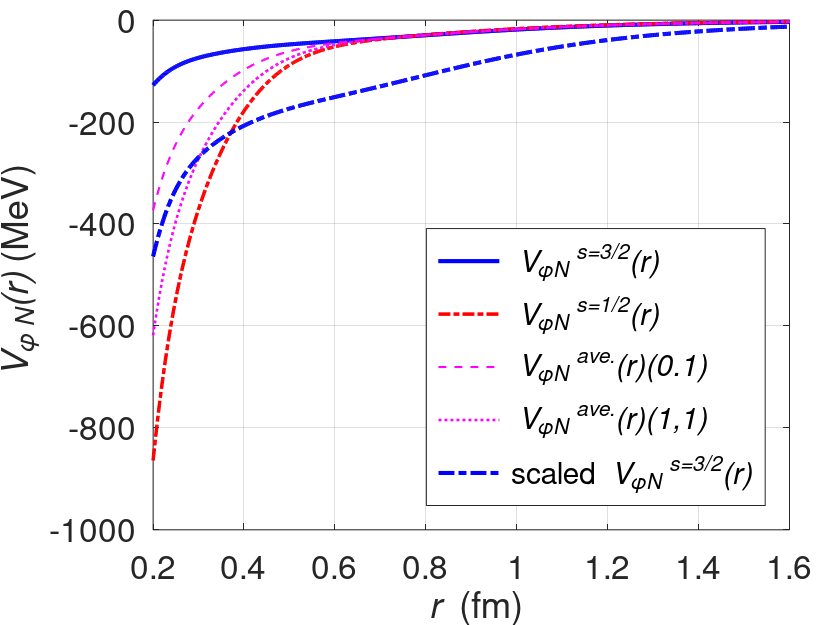}
\end{center}
\caption{ The $\phi N$ potentials for deferent spin-isopin states and modifications according to Table \ref{RtR3}.
} \label{f:1}
\end{figure}

\begin{figure}[ht]
\begin{center}
\includegraphics[width=20pc]{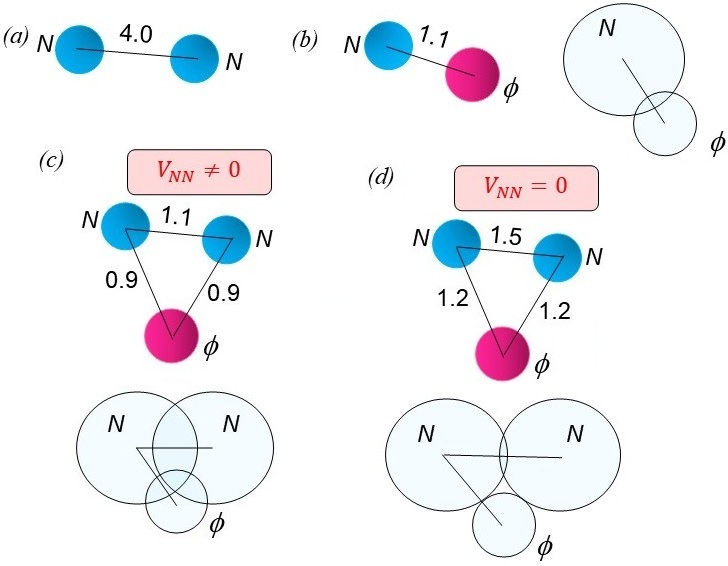}
\end{center} \label{f:7}
\caption{
The graphic illustration depicts the root-mean-square ($rms$) distances, $d_i$, calculated using Eq. (\ref{meand}), between each pair of particles (in fm) in the \(\phi NN\) bound system ($i = 1,2,3$) and its subsystems.
The {$NN$ interaction} is modeled using the {MT-I-III potential} \cite{MTcorr}, while the {$\phi N$ interaction} is described by the {HAL QCD $\phi N$ potential} in the $^{2}S_{1/2}$ channel \cite{Chizzali2024}.
Additionally, we illustrate the same configurations while considering the {$rms$ radii} of the nucleon, $R^N_{{rms}} = 0.8$ fm, and the{$\phi$ meson}, $R^\phi_{{rms}} = 0.4$ fm. These radii are represented by {weakly filled circles} with corresponding diameters.
   ($a$) The {$rms$ distance} between nucleons in the {deuteron}.
   ($b$) The {$rms$ distance} between the {$\phi$ meson} and a nucleon in the \textbf{$\phi N$} system.
   ($c$) The $rms$ distances, $d_i$, in the \textbf{$\phi NN$} system.
   ($d$) The same as ($c$), but when the {$NN$ interaction is omitted}.
 }
\label{f:7}
\end{figure}
\begin{figure}[t]
\begin{center}
\includegraphics[width=22pc]{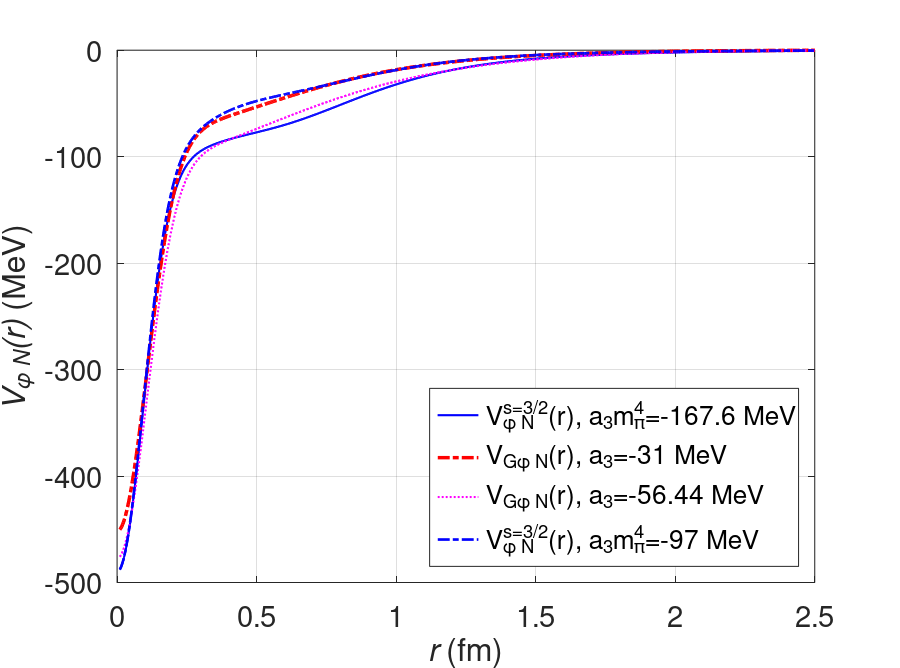}
\end{center}
\caption{ The $V_{\phi N}^{3/2}(r)$ (A-EXP) and $V_{G\phi N}(r)$ (B-3G) model for $\phi N$ potentials from Ref. \cite{EA24} for different parameters
used in Eqs. (\ref{HALQCD}) and (\ref{Gauss3}). Results of calculations for the $\phi N N$ system are presented in Tabl. \ref{Rt33}.
\label{f:2} }
\end{figure}

Calculations similar to those performed in our work were reported in Ref.~\cite{EA24} for the $\phi N$ in the spin 3/2 channel using the potential from Ref.~\cite{Lyu22}, expressed via Eqs.~(\ref{HALQCD}) and (\ref{Gauss3}). The authors identified a bound state in the $\phi NN$ system, which contradicts our findings.  
To reproduce the results of Ref.~\cite{EA24}, we systematically varied the parameters of the potentials. The differences between the original and modified potentials are illustrated in Fig.~\ref{f:2}. The results of our calculations for the corrected potentials are summarized in Table~\ref{Rt33}, alongside a comparison with the results of Ref.~\cite{EA24}. This analysis demonstrates that a $\phi NN$ bound state becomes possible only when the $\phi N$ potential is significantly altered. 
To verify our computational framework, we performed the binding energy calculation  using a Yukawa-type potential motivated by the QCD van der Waals attractive force, mediated by multi-gluon exchanges~\cite{G2001}. These calculations yielded results consistent with those reported in Refs.~\cite{BSS, Bel2008, Sofi}, which were derived using differential Faddeev equations~\cite{BSS} and the theoretical formalism outlined in~\cite{Sofi}. A summary of these results is provided in Table~\ref{Rt33}, and their agreement supports the validity of our approach.  
Notably, the Yukawa-type $\phi N$ potential from Ref.~\cite{G2001} predicts  deep $S=2$ bound states in $\phi N$ and $\phi NN$ systems.

\subsection{Wood-Saxon type potentials for $\phi$-$\alpha$ interaction}
\label{Calculations1}
In this subsection, we follow our works \cite{FilKezalfa24,FilKezOmega24} and present the results of calculations for the feasibility of expected bound states for $^5_{\phi}$He, $^6_{\phi\phi}$He, and $^9_\phi$Be $\phi$-mesic nuclei in the framework of a two- and three-particle cluster models.

At large distances between $\alpha$ and $\phi$ meson, the clustering is described as $\phi +(NNNN)$.
In the region near the alpha cluster and inside, the clustering must include different combinations $(\phi NNN)+N$,
$(\phi NN )+(NN)$, $(\phi N)+(NNN)$.  Taking into account that the $\phi $ meson does not make bound states in the subsystems,
we assume that the $\phi+(NNNN)$ clusterization is dominating. Thus, we assume that  the folding potential is an appropriate approach for the
$\phi$-$\alpha$ interaction including the near and in regions. 

The root-mean-square 
radius is an important and basic property for any composite subatomic system. For $^4$He both the matter and charge $rms$ radii were measured. A matter radius is related to both the proton and the neutron distributions inside a nucleus, whereas nuclear charge radius primarily connected to the proton distribution. The average $rms$ charge radius of $^4$He from electron elastic scattering experiments is 1.681(4) fm \cite{Sick2008}. While 
combined analysis gave the average $rms$ charge radius of  $^4$He  to be 1.6755(28) fm \cite{Marinova}. The result of precise measurements of  $^4$He $rms$ charge radius with the technique of muon-atom spectroscopy gives 1.67824(13) fm \cite{Krauth}. Recently, by examining the near-threshold $\phi$ meson photoproduction data of the LEPS Collaboration \cite{HiraiwaLEPS}, the $rms$ matter radius of $^4$He is measured to be 1.70 ± 0.14 fm \cite{Wang2024}.  From these analyses, the $rms$ charge
radius of $^{4}$He is smaller than the $rms$ matter radius. However, the values of the $rms$ charge and matter radii are within the statistical errors. Whereas, this is an astounding puzzle, in our calculation of the folding potential we used the density that reproduces the $rms$ radii 1.70 ± 0.14 fm \cite{Wang2024}: 1.56 fm, 1.70 fm, and 1.84 fm. The latter allows us to study the influence of the $rms$ on the $\phi$-$\alpha$ potential.  
As follows from Ref. \cite{Wang2024} the simple Gaussian matter distribution model $\rho (r)=\left( \frac{A^{2}}{\pi }\right) ^{3/2}e^{-C^{2}r^{2}}$ gives $\left\langle r^{2}\right\rangle ^{1/2}=\sqrt{3/2C^{2}}$ and describes the experimental data with parameters from \cite{Wang2024}. In Ref. \cite{Thomas97} suggested parametrization of $^4$He density in the form $\rho (r) = B(1+\gamma r{^2})e^{-\lambda r^{2}}$  that reproduces the central depression measured in the charge density and gives $rms = \sqrt{\frac{3}{2}} \sqrt{\frac{5 \alpha +2 \beta }{\beta  (3 \alpha +2 \beta )}}$ =1.56 fm. The distribution of matter density in $^4$ He that leads to the experimental $rms$ within the experimental statistical uncertainty is shown in Fig. \ref{fig:05}.  Let us mention  that the three Gaussian approximation of the HAL QCD potential is very instrumental, which allows to evaluate (\ref{2Dintegral}) for the Gaussian matter 
and  central depression in the charge density distributions in the analytical form, as shown in Appendix A.
\begin{figure}[ht!]
\begin{center}
\includegraphics[width=16pc]{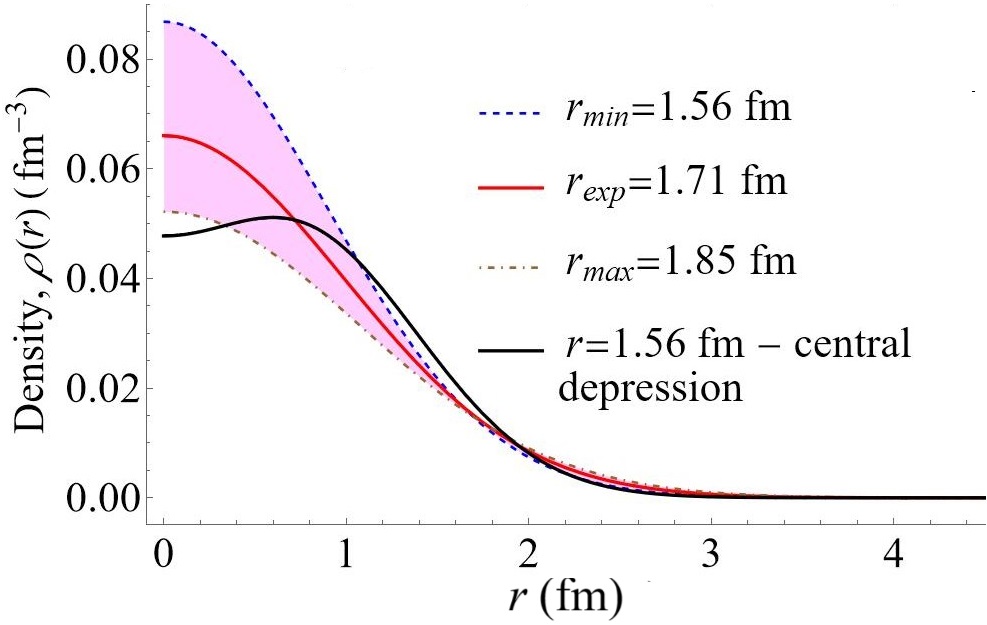}
\end{center}
\caption{The matter density distribution $\rho(r)$ in $^4$He that corresponds to the different $rms$ radii. The red solid curve corresponds to the experimental $rms = 1.70$ fm \cite{Wang2024}. The shaded area corresponds to the experimental uncertainty of $\pm 0.14$ fm for $rms$ radii given by statistics \cite{Wang2024} for the Gaussian distribution. The black solid curve shows the central depression in the charge density \cite{Thomas97}.
}
\label{fig:05}
\end{figure}

For calculations of the binding energies of these systems, we use both the Wood-Saxon (WS) fit for potential simulated from the attractive potential for the $\phi$ meson in the nuclear medium originated
from the in-medium enhanced $K\bar{K}$ loop in the $\phi$ meson self-energy for three values of the cutoff
parameter $\Lambda_{K}$: 2000, 3000, and 4000 MeV \cite{Co17}, denoted as $\widetilde{V}_{\phi\alpha}$ and the WS fit
for the folding 
potential obtained based on the HAL QCD $\phi$-$ N$ potential in the $^{4}S_{3/2}$ channel \cite{Lyu22} denoted as $V_{\phi\alpha}$. 

We utilized the single folding model with the 
mass distribution in $^{4}$He nucleus shown in Fig. \ref{fig:05} and the HAL QCD $V_{\phi N}$ potential to develop a $V_{\phi\alpha}(r)$ potential.
Following \cite{Gal83}, we fit the folding potential using a simple WS type expression
\begin{equation}
V_{\phi \alpha}(r)=-V_{0}\left[ 1+\exp \left( \frac{r-R}{c}\right) \right]^{-1}.
\label{Omegaalfa}
\end{equation}%
In Eq. (\ref{Omegaalfa}) $V_{0}$ is the strength of the interaction, $c$ is the surface diffuseness, and $%
R=1.1A^{1/3}$, where $A=4$ is the mass number of the nuclear core so that $R=1.74 $ fm.
The input parameters for potentials are listed in Table \ref{RKt00}. These potentials are distinguished by the depth of the potential, surface diffuseness, and parameter $R$.

\begin{figure}[ht!]
\begin{center}
\includegraphics[width=20pc]{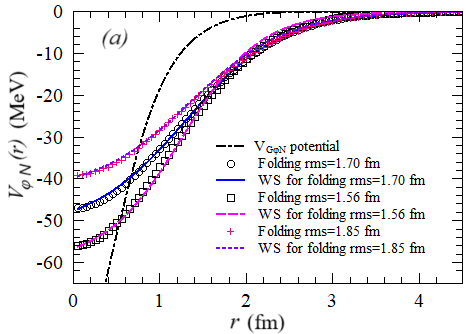}
\includegraphics[width=19pc]{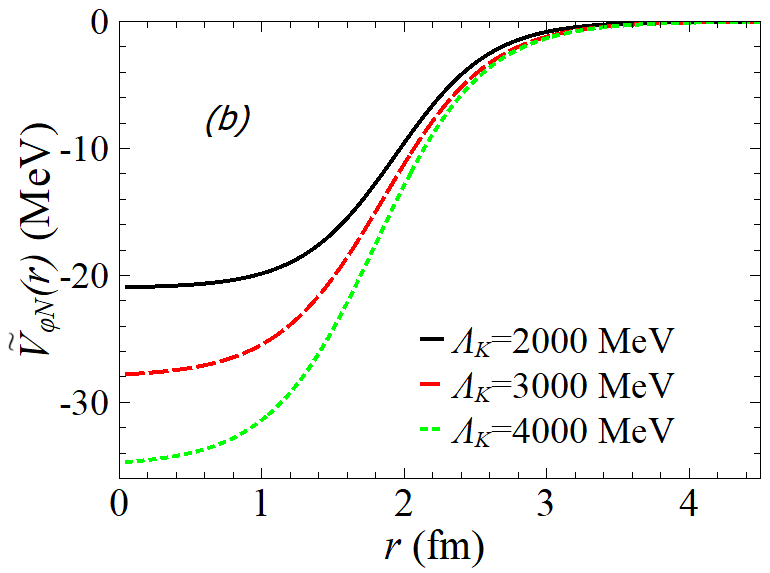}
\end{center}
\caption{($a$) The folding $\phi\alpha$ potentials with the corresponding WS fits. The different symbols correspond to the different $rms$ radii. Calculations were performed for the three values of the $rms$: 1.56 fm, 1.70 fm, and 1.84 fm. The dash-dotted curve depicts the HAL QCD $\phi$-$N$ potential  \cite{Lyu22}. 
($b$) The WS type of $\phi\alpha$ potential simulated from the potential proposed in Ref. \cite{Co17} for three values of the cutoff parameter $\Lambda$: 2000, 300 and 4000 MeV. Figure is adopted from Ref. \cite{FilKezalfa24}.}
\label{fig04}
\end{figure}

First, we present the results of calculations for the folding $\phi$-$\alpha$ potential obtained using the 
density distributions given in Fig. \ref{fig:05}. The WS fit of the folding potential is denoted as $V_{\phi\alpha}$.
The dependence of the folding potential and its WS fit for different $rms$ radii is shown in Fig. \ref{fig04}$a$. One can see that
the depth of the $V_{\phi\alpha}$ potential is very sensitive to the value of
the $rms$ radius, $\it{i.e.}$ the mass density distribution, varying from $\sim -38$ MeV to $-54$ MeV. Let us mention that at the asymptotical region $r>2$ fm the density distribution drops drastically for large distances. The latter leads to the best-fitting parameter  $R=0.856A^{1/3}$ fm for $^4$He.

In Ref. \cite{Thomas97} was chosen the other density model to reproduce the $rms$ matter radius of
$^4$He, 1.56 fm, and the measured central depression in the density. Employing this density matter model the $\phi$ meson–$^4$He potentials were calculated based on an effective
Lagrangian approach \cite{CoPL} using a local density approximation, with the inclusion of the $K\bar{K}$ meson loop in the
$\phi$ meson self-energy \cite{Co17} for different values of a cutoff parameter $\Lambda{_c}$. We simulated this interaction by the WS-type potential. In Fig. \ref {fig04}$b$ depict the fit by (\ref{Omegaalfa}) of the corresponding $\phi$-$^4$He interaction for three values of the cutoff
parameter $\Lambda_{c} = 2000$, 3000, and 4000 MeV. 
The depth of the $\widetilde{V}_{\phi \alpha }$ potential is sensitive to
the cutoff parameter, varying from $\sim -20$ MeV to $-35$ MeV. This is consistence with theoretical and experimental findings that suggest of attractive $\phi$-nuclei potentials with a depth of $\sim-(20-30)$ MeV \cite{Paryev2017}.
The simulation of this $\phi$-$\alpha$ interaction by the WS type potential gives $R=1.24A^{1/3}$ fm, in contrast to $R=0.856A^{1/3}$ obtained in the case when we used the HAL QCD $V_{\phi N}$ potential.

\subsection{ $^5_\phi$He in a two-body $\phi$+$\alpha$ cluster model}
\label{sec:5a}

First, let us consider the $^5_\phi$He $\phi$-mesic nucleus within the two-body cluster model as the $\phi$+$\alpha$ system. Results of calculations for the binding energy and scattering length for the $\phi$+$\alpha$ system with $\widetilde{V}_{\phi \alpha }$ and $V_{\phi \alpha }$ potentials 
are presented in Table \ref{RKt00}. 
The $\widetilde{V}_{\phi \alpha }$ potential has $R$ of about 2 fm in contrast to $R \sim 1.4$ fm for $V_{\phi \alpha }$ but about $\sim 45\%$ weaker 
the interaction strength.  Consequently, the binding energy, $B_{\phi\alpha}$, of the $\phi$+$\alpha$ system ($^{5}_{\phi}$He) obtained using $\widetilde{V}_{\phi \alpha }$ potential is $\sim 0.8 - 5$ MeV, in contrast to $\sim 3 - 6$ MeV utilizing the $V_{\phi \alpha }$ potential.

\begin{table}[ht]
\caption{ Parameters for the WS simulation of the $\widetilde V_{\phi\alpha}$ potential for the $\phi$ meson in the nuclear medium originated
from the in-medium enhanced $K\bar{K}$ loop in the $\phi$ meson self-energy \cite{Co17} for cutoff parameters $\Lambda_c$=2000, 3000, and 4000 MeV.
Parameters of the WS simulation of the $V_{\phi\alpha}$ potential ($V_0$ in MeV, $R$ and $c$ in fm)  generated through a folding procedure from the HAL QCD $\phi N$ potential \cite{Lyu22} employing the Gaussian density function giving different $rms$ radii in fm.
 $B_{\phi\alpha}$,  $B_{\phi\alpha\alpha}$ and $B_{\phi\phi\alpha}$ are two- and three-body BE energies for the $\phi$+$\alpha $, and  $\phi$+$\alpha $+$\alpha $,  $\phi$+$\phi$+$\alpha $ systems, respectively, in MeV. $B_{\phi \phi \alpha}(V_{\phi\phi} = 0)$ is the three-body BE in MeV for the $\phi$+$\phi$+$\alpha $ system when the interaction between two $\phi$ mesons is omitted.
}
\label{RKt00}
\begin{tabular}{ccccccccc}
\hline \noalign{\smallskip}
$\phi \alpha $ potential & $rms$ & $V_{0}$& $R$ & $c$ & $%
B_{\phi\alpha}$ & $B_{\phi \alpha \alpha }$ & $B_{\phi \phi \alpha} $ & $B_{\phi \phi \alpha}(V_{\phi\phi} = 0)$ \\ \hline \noalign{\smallskip}
\ \ \ \ \ \ \ \ \ $\Lambda_c =2000$ MeV & 1.56 & 21 & 1.94 & 0.33 & 0.80 & 3.20 & 1.23 & 1.67\\
 $\widetilde{V}_{\phi \alpha }$ \ \ $\Lambda_c =3000$ MeV & 1.56 & 28 & 1.94 & 0.33 & 3.19 & 6.13 & 5.41 & 6.52 \\
\ \ \ \ \ \ \ \ $\Lambda_c =4000$ MeV & 1.56 & 35 & 1.80 & 0.37 & 4.71 & 9.69& 8.19 & 9.59 \\  \hline
\noalign{\smallskip}
 & 1.85  & 43 & 1.36 & 0.55 & 2.967 & 7.03& 5.01 & 6.06 \\
 $V_{\phi \alpha }$ & 1.70 & 52 & 1.30 & 0.55 & 4.780 & 9.79&  8.32 & 9.72 \\
& 1.56  & 60 & 1.26 & 0.45 & 5.976 &10.86&  10.5 & 12.2 \\ \hline
\end{tabular}
\end{table}

\subsection{ $^9_\phi$Be in a three-body $\phi$+$\alpha$+$\alpha$ cluster model}
\label{sec:5a}

We now present results for the binding energy of the $^{9}_\phi$Be $\phi$-mesic nucleus, calculated using a $\phi + \alpha + \alpha$ cluster model. To predict a possible bound state, we employ an $\alpha$-$\alpha$ potential with orbital angular momentum components $l = 0, 2,$ and $4$ (Refs.~\cite{AliBodmer,FJ}), along with attractive $\widetilde{V}_{\phi \alpha }$ and $V_{\phi \alpha }$ interactions. The $\alpha$-$\alpha$ potential is the Ali-Bodmer type, using version "a" for $l = 0$ and $2$, and version "d" for $l = 4$.
Table \ref{RKt00} presents the calculated bound state energies ($B_{\phi\alpha\alpha}$) for the $^{9}_\phi$Be nucleus. Using the $\widetilde{V}_{\phi \alpha }$ potential with cutoff parameters $\Lambda_{K} = 2000$, 3000, and 4000 MeV (Ref.~\cite{Co17}), we obtain binding energies of 3.30, 6.13, and 9.69 MeV, respectively, for the $\phi + \alpha + \alpha$ system. A stronger $\phi\alpha$ potential, $V_{\phi \alpha }$, derived from the HAL QCD interaction via a folding procedure, yields binding energies in the range of approximately 7 to 11 MeV. These results indicate that the $\phi$-$\alpha$ interaction derived from HAL QCD in the $^{4}S_{3/2}$ channel leads to a more strongly bound $\phi + \alpha + \alpha$ system.

Analysis of data in Table \ref{RKt00} shows that the $V_{\phi\alpha}$ potential has a WS parameter $R \sim 1.3$ fm, which is less than the $rms$ value of 1.7 fm for $^4$He. Meanwhile, the WS parameter for the $\widetilde{V}_{\phi \alpha}$ potential is $R \sim 2$ fm, which is larger than the $rms$  value of 1.7 fm for $^4$He. This suggests that in one case, the $\phi$ meson is placed within the $\alpha$ particle, while in the case of the $\widetilde{V}_{\phi \alpha}$ potential, the $\phi$ meson is distributed throughout the quantum well, not solely within the $\alpha$ particle. Both scenarios are plausible because there is no active effect of the Pauli principle of the $\phi$ meson on nucleons in the $^8$Be or $^4$He core.

\subsection{ $^6_{\phi\phi}$He in the three-body $\phi$-$\phi$-$\alpha$ cluster model} \label{sec:5b}

We calculate a bound state energy of the $^6_{\phi\phi}$He nucleus in the three-body cluster model $\phi$+$\phi$+$\alpha$ using the $\widetilde{V}_{\phi \alpha }$ and $V_{\phi \alpha }$ potentials. In Table \ref{RKt00} we present the BE, $B_{\phi\phi\alpha}$, by employing the stronger $V_{\phi \alpha }$ potential obtained by the folding procedure of the HAL QCD interaction and $\widetilde{V}_{\phi \alpha }$ potential for three values of the cutoff parameter $\Lambda_{c}$. The binding energy, $B_{\phi\phi\alpha}$, is very sensitive to the value of the cutoff parameter $\Lambda_c$ and increases almost 7 times for $\Lambda_c=2000$ and 4000 MeV. Employing the $V_{\phi \alpha }$ potential with the Gaussian density distribution leading to $rms = 1.85$ fm gives $B_{\phi\phi\alpha} = 5.01$ MeV and increases twice for the $rms=1.56$ fm, $B_{\phi\phi\alpha} = 10.5$ MeV.  In Table \ref{RKt00} are given the results of calculations of the $B_{\phi\phi\alpha}(V_{\phi\phi}=0)$, when the interaction between two $\phi$ mesons is neglected and the system is bound due to only $\phi\alpha$ interactions. In this case, the BE of the  $\phi$+$\phi$+$\alpha$ system increases. The increase of the BE is related to the properties of $\phi\phi$ potential (\ref{pp}) constructed in \cite{Bel2008}. The $\phi\phi$ potential (\ref{pp}) combines a very short-range attractive core with $ V_{2} = 1250$ MeV at distances $r < 0.4$ fm and the long-range repulsive  part with $ V_{1} = 1000$ MeV, and effectively is a repulsive. The short-range attraction does not result in a bound state, and the repulsive barrier at distances around 1 fm limits the close approach of the particle pair in the three-body system. Consequently, the attractive part of the potential is cut off, which manifests in the three-body system as an increase in BE when the $\phi$-$\phi$ interaction is omitted.

Notice that both the $\phi$+$\alpha$+$\alpha$ and the $\phi$+$\phi$+$\alpha$ system are more strongly bound with the $V_{\phi \alpha }$ potential generated by the folding HAL QCD $\phi$-$ N$ interaction in the $^{4}S_{3/2}$ channel. The $\widetilde{V}_{\phi \alpha}$ potential yields weaker bound $\phi$+$\alpha$+$\alpha$ and $\phi$+$\phi$+$\alpha$ systems with energies of approximately 3-9 MeV and 1-8 MeV, respectively. The variation in energy depends on the cutoff parameter $\Lambda_K$ \cite{Co17}. However, a comparison between both calculations reveals qualitative agreement between these approaches for a bound state in the $\phi$+$\alpha$+$\alpha$ and $\phi$+$\phi$+$\alpha$ systems.

\subsection{Validity of the cluster approach}
\label{sec:5d}

The description of $^{5}_{\phi}$He, $^{6}_{\phi\phi}$He, and $^{9}_{\phi}$Be $\phi$-mesic nuclei in cluster model is a limited consideration. The best way to justify the validity of the cluster assumption is to compare the results of the proposed theoretical model with experimental data. Unfortunately, today there are no experimental data nor no-core (\textit{ab initio}) calculations conducted for the $^{5}_{\phi}$He, $^{6}_{\phi\phi}$He, and $^{9}_{\phi}$Be nuclei to compare with the presented results. Due to lack of experimental data and \textit{ab initio} calculations for $^{6}_{\phi\phi}$He and $^{9}_{\phi}$Be $\phi$-mesic nucleus, the use of the cluster model is a good starting point, especially if the analysis is explored within the framework of the Faddeev equations, the most accurate approach in few-body physics.  Let us emphasise that, for example, for $^{6}$He nucleus to explore the structure and dynamics of the strongly correlated many-body problem \textit{ab initio} calculation using the no-core shell model \cite{Shirokov2009,Vary2013,Shirokov2018,Navratil2022}, the variational Monte Carlo method \cite{Wiringa2002}, and quantum Monte Carlo method  \cite{QMC2023} were utilized. It was demonstrated in a comprehensive study of many-body correlations and $\alpha$--clustering in the ground-state and low-lying
energy continuum of the Borromean $^{6}$He nucleus \cite{Navratil2018} that it is possible
to reproduce the correct asymptotic behavior of
the $^{6}$He  wave function  in the more limited
approach  \cite{Navratil2013,FilikhinYF2014,Navratil2014}, based solely on the three-cluster $\alpha$+$n$+$n$ model. Although additional short-range six-body correlations are necessary to correctly describe also the interior
of the wave function for both the ground and scattering states \cite{Navratil2018}.

\section{Concluding remarks}
\label{sec:6}

In this paper we studied the possible existence of light $\phi$-mesic nuclei using lattice HAL QCD $\phi N$ interactions.
We evaluated the binding energy of the $\phi NN$ system  in the framework of Faddeev equations in configuration space employing  the HAL QCD $\phi N$ potential in the $^{2}S_{1/2}$ and $^{4}S_{3/2}$ channels with the maximum and minimum spin, respectively. Although the HAL QCD  $\phi N$ potential in the $^{4}S_{3/2}$ channel exhibits attraction, it does not support bound states for either $\phi N$ or $\phi NN$.
Conversely, employing the HAL QCD $\phi N$ potential in the $^{2}S_{1/2}$ channel yields bound states for both $\phi N$ and $\phi NN$. The binding energies of
$\phi N$ and $\phi NN$ are notably sensitive to variations in the enhancement of the short-range attractive part via the factor $\beta$. Considering both potentials, we find binding energies of 4.23 MeV and 6.82 MeV for the states  $S=1$, $T=0$ and $S=1$, $T=1$ (with triplet and singlet  components of the $NN$ MT I-III potential), respectively, when $\beta = 6.9$. 
These $\phi NN$ states mix the spin channels of the $\phi N$ pair. The analysis presented here demonstrates the potential existence of $^{3}_{\phi}\text{H}$. However, it should be noted that the $(2,0)$ state, with the single channel $^{4}\text{S}_{3/2}$, does not yield a bound state. In contrast, the single channel $^{2}\text{S}_{1/2}$ in the $\phi NN$ state $(0,0)$ results in a binding energy of 64.17~MeV.

We investigate the $\phi$-mesic nuclei $^{5}_{\phi}$He, $^{9}_{\phi}$Be, and $^{6}_{\phi\phi}$He within the framework of the cluster model.  Specifically $\phi$+$\alpha$+$\alpha$ and $\phi$+$\phi$+$\alpha$ systems study, using the Faddeev formalism in configuration space. The $\phi\alpha$ potential is contracted through a folding procedure of the HAL QCD $\phi N$ interaction in the $^4S_{3/2}$ channel utilizing different matter distributions of $^4$He. 
The folding procedure is assumed for the single $\phi+ 4N$ channel because there are no open channels 
near or below the $\phi+ 4N$ threshold and obtain the WS fit $V_{\phi\alpha}$ for the 
folding potential. Additionally, we construct a Wood-Saxon type interaction $\widetilde{V}_{\phi \alpha}$ to simulate the $\phi$-$\alpha$ potential, also taken from the literature, based on an effective Lagrangian approach that includes $K\bar{K}$ meson loops in the $\phi$ meson self-energy.

The $V_{\phi\alpha}$ potential leads to strongly bound $\phi$+$\alpha$+$\alpha$ and $\phi$+$\phi$+$\alpha$ systems with BEs of approximately 7-11 MeV and 5-10 MeV, respectively. The variation in energy depends on the matter density distribution in $^4$He that reproduces the matter $rms$ radius of this nucleus within experimental error bars. In calculations, we used the Gaussian matter densities that reproduced the $rms$ radii within the experimental uncertainty.

The $\widetilde{V}_{\phi \alpha}$ potential yields weaker bound $\phi$+$\alpha$+$\alpha$ and $\phi$+$\phi$+$\alpha$ systems with energies of approximately 3-9 MeV and 1-8 MeV, respectively. The variation in energy depends on the cutoff parameter $\Lambda_K$ \cite{Co17}. However, a comparison between both calculations reveals qualitative agreement between these approaches for a bound state in the $\phi$+$\alpha$+$\alpha$ and $\phi$+$\phi$+$\alpha$ systems.

Our calculations demonstrated the feasibility of the existence of the $^{5}_{\phi}$He, $^{9}_{\phi}$Be, and $^{6}_{\phi\phi}$He nuclei with binding energies
in the range of 1-6 MeV, 1-11 MeV and 3-10 MeV, respectively. The range of values of the binding energies relies on the choice of the WS $\phi$-$\alpha$ interaction parameters.

\section*{Acknowledgments}
This work is supported by 
 US National Science Foundation HRD-1345219 award, the DHS (summer research team), and the Department of Energy/National Nuclear Security Administration Award Number DE-NA0004112. R. Ya. K. is supported by the City University of New York PSC CUNY Research Award \# 66109-00 54.

\appendix

\section{Folding potential}
\subsection{Gaussian form for the density distribution}
Substitution of the HAL QCD $V_{\phi N}$ interaction parameterised with
three Gaussian functions (\ref{Gauss3}) and density \cite{Wang2024}
\begin{equation}
\rho (r)=\left( \frac{A^{2}}{\pi }\right) ^{3/2}e^{-A^{2}r^{2}}
\label{RhoGauss}
\end{equation}
in (\ref{2Dintegral}) gives three integrals
\begin{equation}
F(r,u)=a_{i}\int_{0}^{\infty
}dxe^{-A^{2}x^{2}}e^{(x^{2}+r^{2}-2xru)/b_{i}^{2}}x^{2},\text{ \ \ \ \ }%
i=1,2,3.  \label{FirsInt}
\end{equation}%
The integration gives
\begin{equation}
F(r,u)=\frac{a_{i}b_{i}}{4}\frac{e^{-\frac{r^{2}}{b_{i}^{2}}}}{d_{i}^{5}}%
\times \left[ 2b_{i}d_{i}ru+\sqrt{\pi }e^{\frac{r^{2}u^{2}}{%
b_{i}^{2}d_{i}^{2}}}\left( b_{i}^{2}d_{i}^{2}+2r^{2}u^{2}\right) \left(
\text{erf}\left( \frac{ru}{\text{$b_{i}$}d_{i}}\right) +1\right) \right] ,
\label{eqn1}
\end{equation}%
where $d_{i}=\sqrt{1+A^{2}\text{$b_{i}$}^{2}}$.
Using $F(r,u)$ as the integrant in (\ref{2Dintegral}) yields
\begin{eqnarray}
V_{\phi \alpha }^{F}(r) &=&4A^{3}\exp \left( -\frac{r^{2}}{%
b_{1}^{2}+b_{2}^{2}+b_{3}^{2}}\right) \left\{ \frac{a_{1}b_{1}^{3}}{d_{1}^{3}%
}\exp \left[ \left( \frac{1}{b_{2}^{2}}+\frac{1}{b_{3}^{2}}+\frac{1}{%
b_{1}^{2}d_{1}}\right) r^{2}\right] \right.  + \notag \\
&&\frac{a_{2}b_{2}^{3}}{d_{2}^{3}}\exp \left[ \left( \frac{1}{b_{1}^{2}}+%
\frac{1}{b_{3}^{2}}+\frac{1}{b_{2}^{2}d_{2}^{2}}\right) r^{2}\right] +
\notag \\
&&\left. \frac{a_{3}b_{3}^{3}}{d_{3}^{3}}\exp \left[ \left( \frac{1}{%
b_{1}^{2}}+\frac{1}{b_{2}^{2}}+\frac{1}{b_{3}^{2}d_{3}^{2}}\right) r^{2}%
\right] \right\} .
\end{eqnarray}%

\subsection{Central depression in the density}

In Ref. \cite{Thomas97} was chosen to reproduce the $rms$ matter radius of
$^4$He, 1.56 fm, and the measured central depression in charge density the following expression:
\begin{equation}
\rho (r) = B(1+\gamma r{^2})e^{-\lambda r^{2}},
\label{RD}
\end{equation}
where $B = 0.04775$ fm$^{-3}$, $\gamma = 1.34215$ fm$^{-2}$, and $\lambda = 0.904919$ fm$^{-2}$. Substitution of the HAL QCD V$_{G\phi N}$ interaction parameterized with three Gaussian functions (\ref{Gauss3}) and density (\ref{RD})
in (\ref{2Dintegral}) gives three integrals
\begin{equation}
F_{1i}(r)=a_{i}\int_{-1}^{+1}du\int_{0}^{\infty
}dxe^{-\lambda x^{2}}e^{(x^{2}+r^{2}-2xru)/b_{i}^{2}}x^{2},\text{ \ \ \ \ }%
i=1,2,3.  \label{FirsInt1}
\end{equation}
and three integrals
\begin{equation}
F_{2i}(r)=a_{i}\int_{-1}^{+1}du\int_{0}^{\infty
}dxe^{-\lambda x^{2}}e^{(x^{2}+r^{2}-2xru)/b_{i}^{2}}x^{4},\text{ \ \ \ \ }%
i=1,2,3.  \label{FirsInt2}
\end{equation}
The evaluation of the double integrals (\ref{FirsInt1}) and (\ref{FirsInt2}) as partial defined integrals with respect to $x$ that will give functions of $u$ then can be integrated with respect to $u$ and finally obtain
\begin{equation}
F_{1i}(r) = \frac{\sqrt{\pi } a_{i} b_{i}^3  e^{-\frac{\lambda  r^2}{1-b_{i}^2 \lambda}}}{2 \left(1 + b_{i}^2 \lambda\right)^{3/2}}, \text{ \ \ \ \ }%
i=1,2,3.  \label{FirsInt11}
\end{equation}
and
\begin{equation}
F_{2i}(r) = \frac{\sqrt{\pi } b_{i}^3 e^{-\frac{\lambda  r^2}{1 + b_{i}^2 \lambda}} \left(3b_{i}^4 \lambda +3 b_{i}^2+2 r^2\right)}{4 \left(1 + b_{i}^2 \lambda \right)^{7/2}}, \text{ \ \ \ \ }
i=1,2,3.
\label{FirsInt22}
\end{equation}
Consideration of (\ref{FirsInt11}) and (\ref{FirsInt22}) in ({\ref{2Dintegral}) yields
\begin{equation}
V_{\phi \alpha }^{F}(r) = 2\pi ^{3/2} B \sum_{i = 1}^{3} \frac{a_{i} b_i^3}{ c_{i}^{7/2}}e^{-\frac{\lambda }{c_i}r^2} \left(\lambda  (3 \alpha +2 \lambda ) b_i^4+(3 \alpha +4 \lambda ) b_i^2+2 \alpha  r^2+2\right),
\end{equation}
where $c_{i} = 1+\lambda b_{i}^{2}$.


\newpage
\begin{thebibliography}{99}


\bibitem{Sekihara2010}J. Yamagata-Sekihara, D. Cabrera, M. J. V. Vacas, S. Hirenzaki, Formation of $\phi$ Mesic Nuclei.Prog. Theor. Phys. \textbf{124}, 147-162 (2010).

\bibitem{Hayano2010}R. S. Hayano and T. Hatsuda, Hadron properties in the nuclear medium. Rev. Mod. Phys. \textbf{82}, 2949 (2010).

\bibitem{Paryev2017}  V. Metag, M. Nanova, and E. Ya. Paryev,
Meson-nucleus potentials and the search for meson-nucleus bound states. Prog. Part. Nucl. Phys. \textbf{97}, 199 (2017).

\bibitem{Krein2018}G. Krein, A. W. Thomas, and K. Tsushima, Nuclear-bound quarkonia and heavy-flavor hadrons. Prog. Part. Nucl. Phys. \textbf{100}, 161 (2018).

\bibitem{Tolos2020}L. Tolos and L. Fabbietti, Strangeness in Nuclei and Neutron Stars Prog. Part. Nucl. Phys. 112, 103770 (2020)
\bibitem{Proc2024}P. Achenbach, Mini-Proceedings of the "Fourth International Workshop on the Extension Project for the J-PARC Hadron Experimental Facility (HEF-ex 2024)",  arXiv:2409.00366 (2024). 10.48550/arXiv.2409.00366


\bibitem{Gubler2021}P. Gubler Studying the phi meson in nuclear matter by simulating pA reactions in a transport Approach. Few-Body Syst. (2021) {\bf 62}, 53 (2021).

\bibitem{Weise1998}F. Klingl, T. Waas, and W. Weise, Modification of the $\phi$ meson spectrum in nuclear matter. Phys. Lett. B \textbf{431}, 254 (1998).

\bibitem{Oset2001}E. Oset and A. Ramos, Phi decay in nuclei. Nucl. Phys. A \textbf{679}, 616 (2001).

\bibitem{Cabrera2003}D. Cabrera and M. J. Vicente Vacas, Phi meson mass and decay width in nuclear matter. Phys. Rev. C \textbf{67}, 045203 (2003).

\bibitem{Cabrera2017}D. Cabrera, A. N. Hiller Blin, and M. J. Vicente Vacas, $\phi$ meson self-energy in nuclear matter from $\phi N$ resonant interactions' Phys. Rev. C \textbf{95}, 015201 (2017).

\bibitem{QCDF1992}T. Hatsuda and S. H. Lee, QCD sum rules for vector mesons in the nuclear medium. Phys. Rev. C \textbf{46}, R34 (1992).

\bibitem{QCDF1997}F. Klingl, N. Kaiser, and W. Weise, Current correlation functions, QCD sum rules and vector mesons in baryonic matter. Nucl. Phys. A \textbf{624}, 527 (1997).

\bibitem{QCDF2015}P. Gubler and W. Weise, Moments of $\phi$ meson spectral functions in vacuum and nuclear matter. Phys. Lett. B \textbf{751}, 396 (2015).

\bibitem{QCDF2016}P. Gubler and W. Weise, Phi meson spectral moments and QCD condensates in nuclear matter. Nucl. Phys. A \textbf{954}, 125 (2016).

\bibitem{QCDF2022}J. Kim, P. Gubler, and S. H. Lee, $\phi$ meson properties in nuclear matter from QCD sum rules with chirally separated four-quark condensates. Phys. Rev. D \textbf{105}, 114053 (2022).

\bibitem{CoPL}J. J. Cobos-Martínez, K. Tsushima, G. Krein, and A.W. Thomas, $\phi$ meson mass and decay width in nuclear matter and nuclei Phys. Lett. B \textbf{771}, 113 (2017).

\bibitem{Co17} J. J. Cobos-Martínez, K. Tsushima, G. Krein, and A. W. Thomas, $\phi$ meson–nucleus bound states. Phys. Rev. C 96, 035201 (2017).

\bibitem{Gubler2025}L. M. Abreu, P. Gubler, K. P. Khemchandani, A. M. Torres, A. Hosaka, A study of the $\phi N$ correlation function. Phys. Lett. B \textbf{860}, 139175 (2025).

\bibitem{HiyamaXi2020}E. Hiyama, K. Sasaki, T. Miyamoto, T. Doi, T. Hatsuda, Y. Yamamoto, and Th. A. Rijken, Possible lightest $\Xi$ hypernucleus with modern $\Xi N$ interactions, Phys. Rev. Lett. \textbf{124}, 092501 (2020).

\bibitem{Filikhin2000}I. N. Filikhin and S. L. Yakovlev, Calculation of the binding energy and of the parameters of low-energy scattering in the $\Lambda np$ system, Phys. Atom. Nucl., \textbf{63}, 223 (2000).

\bibitem{GV15} H. Garcilazo and A. Valcarce, Light $\Xi$ hypernuclei. Phys. Rev. C \textbf{92}, 014004 (2015).

\bibitem{GV2016}H. Garcilazo and A. Valcarce, Deeply bound $\Xi$ tribaryon, Phys. Rev. C \textbf{93}, 034001 (2016).

\bibitem{GVV16} H. Garcilazo, A. Valcarce, and J. Vijande, Maximal isospin few-body systems of nucleons and $\Xi$ hyperons. Phys. Rev. C  \textbf{94}, 024002 (2016).

\bibitem{FSV17} I. Filikhin, V. M. Suslov and B. Vlahovic, Faddeev calculations for light $\Xi$-hypernuclei, Mat. Model. Geom., \textbf{5}, 1 (2017).

\bibitem{GV0} H. Garcilazo and A. Valcarce, $\Omega d$ bound state, Phys. Rev. C \textbf{98}, 024002 (2018).

\bibitem{GV}  H. Garcilazo and A. Valcarce, $\Omega NN$ and $\Omega \Omega N$ states. Phys. Rev. C \textbf{99}, 014001 (2019).

\bibitem{Gibson2020}B. F. Gibson and I. R. Afnan, Exploring the unknown $\Lambda n$ interaction, SciPost Phys. Proc. \textbf{3}, 025 (2020).


\bibitem{EF21} F. Etminan and M.M. Firoozabadi, $\Omega$-deuteron Interaction in Folding Model, arXiv:1908.11484v5 [nucl-th]

\bibitem{Zhang2022}L. Zhang, Song Zhang, and Y-G. Ma Production of $\Omega NN$ and $\Omega \Omega N$  in ultra-relativistic heavy-ion collisions. Eur. Phys. J. C \textbf{82}, 416 (2022).

\bibitem{GV2022}  H. Garcilazo and A. Valcarce, $(I,J^{P})=(1,1/2^{+})$   $\Sigma NN$ quasibound state, Symmetry \textbf{14}, 2381 (2022).

\bibitem{ESE2023}  F. Etminan, Z. Sanchuli, M. M. Firoozabadi,
Geometrical properties of $\Omega NN$ three-body states by realistic NN and first principles lattice QCD $\Omega N$ potentials Nucl. Phys. A \textbf{1033} 122639 (2023).


\bibitem{Akaishi2002}Y. Akaishi and T. Yamazaki, Phys. Rev. C \textbf{65}, 044005 (2002).

\bibitem{Yamazaki2002} T. Yamazaki and Y. Akaishi, Phys. Lett. B \textbf{535}, 70 (2002).

\bibitem{Yamazaki2004} T. Yamazaki, A. Dote and Y. Akaishi, Phys. Lett. B \textbf{587}, 167 (2004).




\bibitem{Shevchenko2017} N. V. Shevchenko, Three-Body Antikaon–Nucleon Systems, Few-Body Syst 58: 1–25 (2017).

\bibitem{RKezNNNK} R. Ya. Kezerashvili, S. M. Tsiklauri, and N. Zh.
Takibayev, Search and research of $\bar{K}NNN$ and $\bar{K}\bar{K}NN$
antikaonic clusters. Prog. Part. Nucl. Phys. {\bf 121}, 103909 (2021).


\bibitem{BSS} V. B. Belyaev, W. Sandhas, and I. I. Shlyk, 3- and 4- body meson- nuclear clusters, ArXiv: nucl-th0903.1703

\bibitem{Bel2008} V. B. Belyaev, W. Sandhas, and I. I. Shlyk, New nuclear three-body clusters $\phi NN$, Few Body Syst. \textbf{44} 347. (2008).

\bibitem{Sofi} S. A. Sofianos, G. J. Rampho, M. Braun and R. M. Adam, The $\phi$-$NN$ and $\phi\phi$–$NN$ mesic nuclear systems, J. Phys. G: Nucl. Part. Phys. \textbf{37}, 085109 (2010).

\bibitem{EA24} F. Etminan, A. Aalimi, Examination of the $\phi$-$NN$ bound-state problem with lattice QCD $N$-$\phi$ potentials, Phys. Rev. C  \textbf{109}, 054002 (2024).

\bibitem{FKVPRD2024}I. Filikhin, R. Ya. Kezerashvili, and B. Vlahovic, Possible $^{3}_{\phi}$H hypernucleus with HAL QCD interaction. Phys. Rev. D  \textbf{110}, L031502 (2024).


\bibitem{ALICE2021} S. Acharya et al. [ALICE], Experimental evidence for an
attractive $p-\phi $ interaction. Phys. Rev. Lett. \textbf{127}, 172301
(2021).

\bibitem{Lyu22} Y. Lyu, T. Doi, T. Hatsuda, Y. Ikeda, J. Meng, K. Sasaki, and T. Sugiura, Attractive $N$-$\phi$ interaction and two-pion tail from lattice QCD near physical point. Phys. Rev. D {\bf 106}, 074507 (2022).

\bibitem{Chizzali2024}E. Chizzali, Y. Kamiya, R. Del Grande, T.i Doi, L. Fabbietti, T. Hatsuda, and Y. Lyu, Indication of a $p$-$\phi$ bound state from a correlation function analysis, Phys. Lett. B \textbf{848}, 138358 (2024).

\bibitem{Shell1}A. Gal, J. Soper, and R. Dalitz, A shell-model analysis of $\Lambda$ binding energies for the $p$-shell hypernuclei. I. Basic formulas and matrix elements for $\Lambda N$  and $\Lambda NN$  forces. Ann. Phys. (N.Y.) \textbf{63}, 53 (1971).

\bibitem{Shell2}A. Gal, J. Soper, and R. Dalitz, A shell-model analysis of $\Lambda$ binding energies for the $p$-shell hypernuclei II. Numerical Fitting. Interpretation, and Hypernuclear Predictions \textbf{72}, 445 (1972).

\bibitem{Shell3}A. Gal, J. Soper, and R. Dalitz, A shell-model analysis of $\Lambda$ binding energies for the $p$-shell hypernuclei III. Further analysis and predictions. Ann. Phys. (N.Y.)  \textbf{113}, 79 (1978).

\bibitem{MShell1} D. J. Millener, Shell-model description of $\Lambda$ hypernuclei. Nucl. Phys. A \textbf{691}, 93 (2001).

\bibitem{MShell2} D. J. Millener, hell-model interpretation of $\gamma$-ray transitions in p-shell hypernuclei. Nucl. Phys. A \textbf{804}, 84 (2008).
\bibitem{MShell3}D. J. Millener, Shell-model structure of light hypernuclei. Nucl. Phys. A \textbf{835}, 11 (2010).
\bibitem{MShell4}D. J. Millener, Shell-model calculations for p-shell hypernuclei. Nucl. Phys. A \textbf{881}, 298 (2012).

\bibitem{NoShell1}R. Wirth, D. Gazda, P. Navrátil, A. Calci, J. Langhammer, and R. Roth, Ab initio description of $p$-shell hypernuclei. Phys. Rev. Lett. \textbf{113}, 192502 (2014).

\bibitem{MShell5}A. Gal, E. V. Hungerford, D. J. Millener, Strangeness in nuclear physics, Rev. Mod. Phys. \textbf{88}, 035004 (2016).


\bibitem{Wirth} R.  Wirth, R. Roth, Similarity renormalization group evolution of
hypernuclear Hamiltonians. Phys. Rev. C {\bf 100}, 044313 (2019);
R. Wirth, D. Gazda, P. Navratil, and R. Roth, Hypernuclear no-core shell model. Phys. Rev.  C {\bf 97}, 064315 (2018).

\bibitem{MShell6} H. Le, J. Haidenbauer, U.-G. Meisner, and A. Nogga, Jacobi no-core shell model for $p$-shell hypernuclei. Eur. Phys. J. A \textbf{56}, 301 (2020).

\bibitem{MShell7}D. Gazda, T. Yadanar Htun, and C. Forssén, Nuclear physics uncertainties in light hypernuclei. Phys. Rev. C \textbf{106}, 054001 (2022).

\bibitem{Hiyama16}P. Vesel\'{y}, E. Hiyama, J. Hrt\'{a}nkov\'{a}, J. Mare\v{s}, Sensitivity of $\Lambda$ single-particle energies to the $\Lambda NN$ spin–orbit coupling and to nuclear core structure in $p$-shell and $sd$-shell hypernuclei. Nucl. Phys. A \textbf{954}, 260 (2016).


\bibitem{Motoba832} H. Bando, K. Ikeda, and T. Motoba, Coupling features in $^9_\Lambda$Be, $^{13}_\Lambda$C and $^{21}_\Lambda$Ne hypernuclei  Prog. Theor. Phys. \textbf{69}, 918 (1983).

\bibitem{Motoba83} T. Motoba, H. Bando, and K. Ikeda, Light p-shell $\Lambda$-hypernuclei by the microscopic three-cluster model. Prog. Theor. Phys. \textbf{70}, 189 (1983).


\bibitem{Motoba85} T. Motoba, H. Bando, K. Ikeda, and T. Yamada, Production, structure and decay of light p-shell $\Lambda$-hypernuclei. Prog. Theor. Phys. Suppl. \textbf{81}, 42 (1985).

\bibitem{HiyamaCluster2001}E. Hiyama, M. Kamimura, T. Motoba, T. Yamada, and Y. Yamamoto, Three- and four-body structure of light hypernuclei. Nucl. Phys. A \textbf{684}, 227 (2001).

\bibitem{HiyamaPPNP2009} E. Hiyama and T. Yamada, Structure of light hypernuclei. Prog. Part. Nucl. Phys. \textbf{63}, 339 (2009).

\bibitem{Okada2024}M. Okada, W. Horiuchi,  and N. Itagaki, Shell-cluster transition in $^{48}$Ti. Phys. Rev. C \textbf{109}, 054324 (2024).

\bibitem{Clark2024}J. W. Clark and E. Krotscheck, $\alpha$-cluster matter reexamined. Phys. Rev.C \textbf{109}, 034315 (2024).


\bibitem{Fujiwara2004} Y. Fujiwara, K. Miyagawa, M. Kohno, Y. Suzuki, D.
Baye, J.-M. Sparenberg, Faddeev calculation of 3$\alpha $ and $\alpha \alpha
\Lambda $ systems using resonating-group method kernels,
Phys. Rev. C \textbf{70}, 024002 (2004).

\bibitem{Suslov2004} V. M. Suslov, I. Filikhin and B Vlahovic, Cluster
calculation for $_{\Lambda }^{9}$Be hypernucleus. J. Phys. G \textbf{30},
513 (2004)

\bibitem{IgorF2004} I. Filikhin, A. Gal, and V. M. Suslov, Cluster models of
$_{\Lambda \Lambda }^{6}$He and $_{\Lambda }^{9}$Be hypernuclei. Nucl. Phys.
A \textbf{743}, 194 (2004).

\bibitem{Hiyama2012FBS} E. Hiyama, Few-body aspects of hypernuclear physics.
Few-Body Syst \textbf{53}, 189 (2012).

\bibitem{Wu2020} Q. Wu, Y. Funaki, E. Hiyama, and H. Zong, Resonant states
of $_{\Lambda }^{9}$Be with $\alpha +\alpha +\Lambda $ three-body cluster
model. Phys. Rev. C \textbf{102}, 054303 (2020).

\bibitem{HiyamaRev2018} E. Hiyama and K. Nakazawa, Structure of $S=-2$
hypernuclei and hyperon--hyperon interactions. Annu. Rev. Nucl. Part. Sci.
\textbf{68}, 131 (2018).
\bibitem{Lazauskas2020} R. Lazauskas, J. Carbonell, Description of Four- and Five-Nucleon Systems by Solving Faddeev-Yakubovsky Equations in Configuration Space, Frontiers in Physics \textbf{7} 251 (2020).

\bibitem{Strakovsky2020} I. I. Strakovsky, L. Pentchev and A. Titov,
Comparative analysis of $\omega p,$ $\phi p$ and $J/\psi p$ scattering
lengths from A2, CLAS, and GlueX threshold measurements, Phys. Rev. C
\textbf{101}, 045201 (2020).

\bibitem{Brodsky1990}S. J. Brodsky, I. A. Schmidt, and G. F. de Teramond, Nuclear-bound quarkonium, Phys. Rev. Lett. \textbf{64}, 1011 (1990).

\bibitem{G2001} H. Gao,  T.-S. H. Lee, V. Marinov, $\varphi - N$ bound state. Phys. Rev. C \textbf{63}, 022201 (2001).


\bibitem{Sasaki2020}K.Sasaki, et al., 
(HAL QCD Collaboration)$\Lambda\Lambda$ and $N\Xi$ interactions from lattice QCD near the physical point. Nucl. Phys. A
\textbf{998}, 121737 (2020).


\bibitem{Iritani2019} T. Iritani et al., $N \Omega$ dibaryon from lattice QCD near the physical point. Phys. Lett. B \textbf{792}, 284
(2019).

\bibitem{OmegaOmega2018} S. Gongyo, et al., Most strange dibaryon from
lattice QCD, Phys Rev. Lett. \textbf{120}, 212001 (2018).


\bibitem{Kreinm4}J. Tarrús Castell\`{a} and G. A. Krein, Effective field theory for the nucleon-quarkonium interaction. Phys. Rev. D \textbf{98}, 014029
(2018).

\bibitem{Wiringa95}R. B. Wiringa, V. G. J. Stoks, and R. Schiavilla, Accurate nucleon-nucleon potential with charge-independence breaking.  Phys. Rev. C \textbf{51}, 38 (1995).

\bibitem{Malfliet1969}R. Malfliet and J. Tjon, Solution of the Faddeev equations for the triton problem using local two-particle interactions, Nucl. Phys. A \textbf{127} 161–168 (1969). https://doi .org /10 .1016 /0375 -9474(69 )90775 -1, https://www.sciencedirect .com /science /article /pii /0375947469907751.

\bibitem{MTcorr}J. L. Friar, B. F. Gibson, G. Berthold, W. Glockle,
Th. Cornelius, H. Witala, J. Haidenbauer, Y. Koike,
G. L. Payne, J. A. Tjon, and W. M. Kloet, Benchmark solutions for a model three-nucleon scattering problem, Phys. Rev.
C \textbf{42}, 1838 (1990).

\bibitem{Satchler1983}G. R. Satchler, Direct Nuclear Reactions, (Oxford University Press, New York, 1983).

\bibitem{AliBodmer} S. Ali and A. R. Bodmer, Phenomenological $\alpha$-$\alpha$ potentials. Nucl. Phys. \textbf{80},  99 (1966).
\bibitem{FJ} D.V. Fedorov, AS. Jensen The three-body continuum Coulomb problem and the 3$\alpha$ structure of $^{12}$C, Physics Letters B \textbf{389},  631  (1996).

\bibitem{Morse1929} P. M. Morse, Diatomic Molecules According to the Wave Mechanics. II. Vibrational Levels. Phys. Rev. \textbf{34}, 57 (1929).

\bibitem{Hulthen2016} J. Bhoi and U. Laha, Elastic scattering of light nuclei through a simple potential model. Phys. At. Nucl. \textbf{79}, 370 (2016).

\bibitem{Jibuti1978}R. I. Jibuti, N. B. Krupennikova, and V. Yu. Tomchinski, Yad. Fiz. \textbf{28}, 30 (1978).

\bibitem{KezNP1984}R. I. Jibuti and R. Ya. Kezerashvili, Quasi-$\alpha$-particle mechanism of pion double charge exchange on light nuclei, Nucl. Phys A. \textbf{430}, 573 (1984).

\bibitem{KezYad1992}R. I. Dzhibuti, R. Ya. Kezerashvili, and N. I. Shubitidze, Photodisintegration of the $\alpha$-cluster nuclei into the $\alpha$-particles, Yad. Fiz.  \textbf{55}, 3233 (1992); Sov. J. Nucl. Phys. \textbf{55}, 1801 (1992).

\bibitem{Igor2000}I. N. Filikhin, $\alpha$+$^8$Be cluster model for $0+2$ resonance in the $^{12}$C nucleus, Phys. Atom. Nucl. \textbf{63}, 1527 (2000).

\bibitem{fed1}D. V. Fedorov, E. Garrido, A. S. Jensen, Complex Scaling of the Hyper-Spheric Coordinates and Faddeev Equations. Few Body Syst. \textbf{33}, 153 (2003).

 \bibitem{suslov2004} V. M. Suslov, I. Filikhin, and B. Vlahovic, Cluster calculation for $^{9}_
\Lambda$Be hypernucleus, J. Phys. G \textbf{30}, 513 (2004).


 \bibitem{suslov2005}I. Filikhin, V. M. Suslov, and B. Vlahovic, 0$^{+}$ states of the $^{12}$C nucleus: the Faddeev calculation in
configuration space. J. Phys. G: Nucl. Part. Phys. \textbf{31}, 1207 (2005).

\bibitem{fed2}R. Alvarez-Rodriguez, A. S. Jensen, D. V. Fedorov, H. O. U. Fynbo, E. Garrido, Structure of low-lying $^{12}$C resonances. Eur. Phys. J. A \textbf{ 31}, 303 (2007).

\bibitem{fed3}R. Alvarez-Rodriguez, A. S. Jensen, E. Garrido, D. V. Fedorov, H. O. U. Fynbo, alpha particle momentum distributions from C-12 decaying resonances. Phys. Rev. C \textbf{77}, 064305 (2008).


\bibitem{Ishikawa1}S. Ishikawa, Three-body calculations of the triple-$\alpha$ reaction. Phys. Rev. C \textbf{87}, 055804 (2013). 

\bibitem{Ishikawa2}S. Ishikawa, Decay and structure of the Hoyle state. Phys. Rev. C \textbf{90}, 061604(R) (2014).

\bibitem{Ishikawa3}S. Ishikawa, Structure of resonance states in three-alpha systems. Few-Body Syst. \textbf{65}, 50 (2024).


\bibitem{Fad} L. D. Faddeev, Scattering theory for a three-particle system.
ZhETF \textbf{39}, 1459 (1961); [Sov. Phys. JETP \textbf{12}, 1014 (1961)].

\bibitem{Fad1} L. D. Faddeev, Mathematical problems of the quantum theory of
scattering for a system of three particles. Proc. Math. Inst. Acad. Sciences
USSR \textbf{69}, 1 (1963).

\bibitem{Noyes1968} H. P. Noyes and H. Fiedeldey, In: Three-Particle
Scattering in Quantum Mechanics (Gillespie, J., Nutall, J., eds.), p. 195.
New York, Benjamin, 1968.

\bibitem{Noyes1969} H. P. Noyes In: Three-Body Problem in Nuclear and
Particle Physics (\textit{Proceedings of the 1st Int. Conf., Birmingham, 1969%
}), (McKee, J. S. C., Rolph, P. M., eds.), p. 2. Amsterdam, North-Holland,
1970.

\bibitem{Gignoux1974} C. Gignoux, C., Laverne, and S. P. Merkuriev, Solution
of the Three-Body Scattering Problem in Configuration Space, Phys. Rev.
Lett. {\bf 33}, 1350 (1974).

\bibitem{FM} L.D. Faddeev and S.P. Merkuriev, {Quantum Scattering Theory
for Several Particle Systems} (Kluwer Academic, Dordrecht, 1993) pp. 398.

\bibitem{K86} A.A. Kvitsinsky, Yu.A. Kuperin, S.P. Merkuriev, A.K. Motovilov
and S.L. Yakovlev, N-body Quantum Problem in Configuration Space. \textit{%
Fiz. Elem. Chastits At. Yadra} \textbf{17}, 267 (1986) (in Russian);\newline
http://www1.jinr.ru/Archive/Pepan/1986-v17/v-17-2.htm




\bibitem{Kez2017} R. Ya. Kezerashvili, Sh. M. Tsiklauri, I. Filikhin, V. M.Suslov, B. Vlahovic, Three-body calculations for the $K^{-}pp$ system within potential models. J. Phys. G: Nucl. Part. Phys. \textbf{43}, 065104 (2016).

\bibitem{Kez2018PL} I. Filikhin, R. Ya. Kezerashvili, and B. Vlahovic, On binding energy of trions in bulk materials. Phys.
Lett. A \textbf{382}, 787 (2018).


\bibitem{KezPRD2020}I. Filikhin, R. Ya. Kezerashvili, V. M. Suslov, Sh. M. Tsiklauri, and B. Vlahovic, Three-body model for $K(1460)$ resonance.  Phys. Rev. D \textbf{102}, 094027 (2020).

\bibitem{KezJPG2024}I. Filikhin, R. Ya. Kezerashvili, and B. Vlahovic, The charge and mass symmetry breaking in
the $KK\bar{K}$ system. J. Phys. G: Nucl. Part. Phys. \textbf{51}, 035102 (2024).



\bibitem{H2002} E. Hiyama, M. Kamimura, T. Motoba, T. Yamada and Y.
Yamamoto, Four-body cluster structure of A=7-10 double-$\Lambda$ hypernuclei. Phys. Rev. C \textbf{66}, 024007-13 (2002).

\bibitem{FG2002} I.N. Filikhin and A. Gal, Light $\Lambda\Lambda$ hypernuclei and the onset of stability for $\Lambda\Xi$ hypernuclei. Phys. Rev. C \textbf{65},
041001(R)-4 (2002).

\bibitem{F2018} I. Filikhin, R. Ya. Kezerashvili, V. M. Suslov and B. Vlahovic, On mass polarization effect in three-body systems, Few-Body Syst. \textbf{59}, 33 (2018).

\bibitem{FilKezalfa24}I. Filikhin, R. Ya. Kezerashvili, and B. Vlahovic, Bound states of $^9_\phi$Be and $^6_{\phi\phi}$He within $\phi$ + $\alpha$ +$\alpha$ and $\phi$ + $\phi$ + $\alpha$ cluster models. Phys. Rev. C \textbf{110}, 065202 (2024).

\bibitem{FilKezOmega24}I. Filikhin, R. Ya. Kezerashvili, and B. Vlahovic, 
Folding procedure for $\Omega-\alpha$ potential. Few-Body Syst. \textbf{66}, 2 (2025).



\bibitem{Sick2008}I. Sick, Precise root-mean-square radius of $^4$He. Phys. Rev. C {\bf 77}, 041302(R) (2008).

\bibitem{Marinova}I. Angeli and K. P. Marinova, Table of experimental nuclear ground state charge radii: An update. At. Data Nucl. Data Tables {\bf 99}, 69 (2013).

\bibitem{Krauth}J. J. Krauth et al., Measuring the $\alpha$-particle charge radius with muonic helium-4 ions. Nature (London) {\bf 589}, 527 (2021).

\bibitem{HiraiwaLEPS}T. Hiraiwa et al. (LEPS Collaboration), First measurement of coherent $\phi$ meson photoproduction from $^4$He near threshold. Phys. Rev. C {\bf 97}, 035208 (2018).

\bibitem{Wang2024}R. Wang, C.Han, and X. Chen, Exploring the mass radius of $^{4}$He and implications for nuclear structure. Phys. Rev. C {\bf 109}, L012201 (2024).



\bibitem{Thomas97}K. Saito, K. Tsushima, and A.W. Thomas, $\rho$-meson mass in light nuclei. Phys. Rev. C \textbf{56}, 566 (1997).

\bibitem{Gal83}C. B. Dover and A. Gal, $\Xi$ Hypernuclei. Ann. Phys. \textbf{146}, 309 (1983).

\bibitem{Navratil2022}A. Glick-Magid, C. Forssén, D. Gazda, D. Gazit, P. Gysbers, and P. Navr\'atil, Nuclear \textit{ab initio} calculations of $^{6}$He $\beta$-decay for beyond the Standard Model studies.
Phys. Lett. B \textbf{832}, 137259 (2022).


\bibitem{Shirokov2009} P. Maris, J. P. Vary, and A. M. Shirokov, Ab initio no-core full configuration calculations of light nuclei, Phys. Rev. C \textbf{79}, 014308 (2009).


\bibitem{Vary2013} B. R. Barrett, P. Navr\'atil, J. P. Vary, Ab initio no core shell model, Prog. Part. Nucl. Phys. \textbf{69}, 131 (2013).

\bibitem{Shirokov2018} A. M. Shirokov, A. I. Mazur, I. A. Mazur, E. A. Mazur, I. J. Shin, Nucleon-$\alpha$  Scattering and Resonances in $^5$He and $^5$Li with JISP16 and Daejeon16 $NN$ interactions,  Phys. Rev. C \textbf{98}, 044624 (2018).

\bibitem{Wiringa2002}R. Schiavilla and R. B. Wiringa, Weak transitions in $A=6$ and 7 nuclei. Phys. Rev. C \textbf{65}, 054302 (2002).


\bibitem{QMC2023}G. B. King, A. Baroni, V. Cirigliano, S. Gandolfi, L. Hayen, E. Mereghetti, S. Pastore, and M. Piarulli, \textit{Ab initio} calculation of the $\beta$-decay spectrum of $^{6}$He. Phys. Rev.  \textbf{107}, 015503 (2023).

\bibitem{Navratil2018}S. Quaglioni, C. Romero-Redondo, P. Navr\'atil, and G. Hupin, Three-cluster dynamics within the \textit{ab initio} no-core shell model with continuum:
How many-body correlations and $\alpha$-clustering shape $^{6}$He. Phys. Rev. C \textbf{97}, 034332 (2018).

\bibitem{Navratil2013}S. Quaglioni, C. Romero-Redondo, and P. Navr\'atil, Phys. Rev. C \textbf{88}, 034320 (2013).

\bibitem{FilikhinYF2014}I. Filikhin, V. M. Suslov, and B. Vlahovic, Cluster calculation for $^{6}$He spectrum. Phys. Atom. Nucl, \textbf{77}, 384 (2014).


\bibitem{Aziz1979}R. A. Aziz, V. P. S. Nain, J. C. Carley, W. J. Taylor, and G. T. McConville, An accurate intermolecular potential for helium. J. Chem. Phys. \textbf{70}, 4330 (1979).


\bibitem{Navratil2014}C. Romero-Redondo, S. Quaglioni, P. Navr\'atil, and G. Hupin, Phys. Rev. Lett. \textbf{113}, 032503 (2014).



\end{thebibliography}


\end{document}